\documentclass[11pt]{article}
\usepackage{amsmath,amssymb}
\makeatletter
\@addtoreset{equation}{section}
\makeatother

\def\crampest{\medmuskip = 1mu plus 1mu minus 1mu}

\def\ben{\begin{equation}}
\def\een{\end{equation}}

 \let\m=\mu \let\n=\nu  \let\p=\pi

\let\C=\Chi

\def\nn{\nonumber} \def\bd{\begin{document}} \def\ed{\end{document}}
\def\ds{\documentstyle} \let\fr=\frac \let\bl=\bigl \let\br=\bigr
\let\Br=\Bigr \let\Bl=\Bigl
\let\bm=\bibitem
\let\na=\nabla
\let\pa=\partial \let\ov=\overline
\newcommand{\be}{\begin{equation}}
\newcommand{\ee}{\end{equation}}
\def\ba{\begin{array}}
\def\ea{\end{array}}
\def\ft#1#2{{\textstyle{\frac{\scriptstyle #1}{\scriptstyle #2} } }}
\def\fft#1#2{{\frac{#1}{#2}}}
\def\del{\partial}
\def\vp{\varphi}
\def\sst#1{{\scriptscriptstyle #1}}
\def\oneone{\rlap 1\mkern4mu{\rm l}}
\def\td{\tilde}
\def\wtd{\widetilde}
\def\ie{{\it i.e.\ }}
\def\dalemb#1#2{{\vbox{\hrule height .#2pt
        \hbox{\vrule width.#2pt height#1pt \kern#1pt
                \vrule width.#2pt}
        \hrule height.#2pt}}}
\def\square{\mathord{\dalemb{6.8}{7}\hbox{\hskip1pt}}}
\newcommand{\ho}[1]{$\, ^{#1}$}
\newcommand{\hoch}[1]{$\, ^{#1}$}
\newcommand{\bea}{\setlength\arraycolsep{2pt} \begin{eqnarray}}
\newcommand{\eea}{\end{eqnarray}}
\newcommand{\ra}{\rightarrow}
\newcommand{\lra}{\longrightarrow}
\newcommand{\Lra}{\Leftrightarrow}
\newcommand{\bp}{\tilde \beta^\prime}
\newcommand{\tr}{{\rm tr} }
\newcommand{\Tr}{{\rm Tr} }
\def\0{{\sst{(0)}}}
\def\1{{\sst{(1)}}}
\def\2{{\sst{(2)}}}
\def\3{{\sst{(3)}}}
\def\4{{\sst{(4)}}}
\def\5{{\sst{(5)}}}
\def\6{{\sst{(6)}}}
\def\7{{\sst{(7)}}}
\def\8{{\sst{(8)}}}
\def\m{{\sst{(m)}}}
\def\n{{\sst{(n)}}}
\def\cA{{{\cal A}}}
\def\cB{{{\cal B}}}
\def\cF{{{\cal F}}}
\def\cG{{{\cal G}}}
\def\cH{{{\cal H}}}
\def\tV{\widetilde V}
\def\tW{\widetilde W}
\def\tH{\widetilde H}
\def\tE{\widetilde E}
\def\tF{\widetilde F}
\def\tA{\widetilde A}
\def\im{{{\rm i}}}
\def\tY{{{\wtd Y}}}
\def\ep{{\epsilon}}
\def\vep{{\varepsilon}}
\def\bD{{{\bar D}}}
\def\R{{{\mathbb R}}}
\def\C{{{\mathbb C}}}
\def\H{{{\mathbb H}}}
\def\CP{{{\mathbb C}{\mathbb P}}}
\def\RP{{{\mathbb R}{\mathbb P}}}
\def\Z{{{\mathbb Z}}}
\def\bA{{{\mathbb A}}}
\def\bB{{{\mathbb B}}}
\def\bC{{{\mathbb C}}}
\def\bD{{{\mathbb D}}}
\def\bE{{{\mathbb E}}}
\def\bZ{{{\mathbb Z}}}
\def\Re{{{\frak{Re}}}}
\def\Im{{{\frak{Im}}}}
\def\cosec{{\,\hbox{cosec}\,}}
\def\Gm{{\Gamma_{\!\! -}}}
\def\Gp{{\Gamma_{\!\! +}}}
\def\stan{{standard }}
\def\nonstan{{supernumerary }}
\def\p{{\partial}}
\def\kdel#1{{\fft{\del}{\del#1}}}

\def\bog{{Bogomolny }}
\def\om{{\omega}}

\newcommand{\nnr}{\nonumber \\}
\newcommand{\pd}{\partial}
\newcommand{\ud}{\textrm{d}}
\newcommand{\ue}{\textrm{e}}
\newcommand{\dTH}{T^{\prime \, 0}_\textrm{H}}
\newcommand{\dOi}{\Omega^{\prime \, 0}_i}

\newcommand{\tamphys}{\it George and Cynthia Woods Mitchell Institute
for Fundamental Physics and Astronomy,\\
Texas A\&M University, College Station, TX 77843-4242, USA}

\newcommand{\auth}{
David D.K. Chow\hoch{\dagger}, M. Cveti\v c\hoch{\$},
H. L\"u\hoch{\dagger\star}, and C.N. Pope\hoch{\dagger,\ddagger}
}

\begin{document}

\begin{flushright}
\hfill{
MIFP-08-34\ \ \ \ \ \ \ \  
UPR-1204-T}\\
 %\hfill{
%\bf hep-th/yymmnnn}
\end{flushright}

%\vspace{25pt}
\begin{center}
{\large {\bf Extremal Black Hole/CFT Correspondence in
(Gauged) Supergravities}}

\vspace{15pt}
\auth

\vspace{10pt}
\hoch{\dagger}{\tamphys}

\vspace{10pt}

\hoch{\$}{\it Department of Physics and Astronomy,\\
University of Pennsylvania, Philadelphia, PA 19104-6396, USA}

\vspace{10pt}

%\hoch{\ddagger}{\it Interdisciplinary Center of Theoretical Studies,
%USTC, Hefei, Anhui 230026, PRC}

\hoch{\star}{\it Division of Applied Mathematics and Theoretical
Physics,\\
China Institute for Advanced Study,\\
Central University of Finance and Economics, Beijing, 100081, China
}

\vspace{10pt}

\hoch{\ddagger}{\it  DAMTP, Centre for Mathematical Sciences,
 University of Cambridge,\\  Wilberforce Road, Cambridge CB3 OWA, UK}

\vspace{30pt}

\underline{ABSTRACT}
\end{center}

   We extend the investigation of the recently proposed Kerr/CFT
correspondence to large classes of rotating black hole solutions in
gauged and ungauged supergravities.  The correspondence, proposed
originally for four-dimensional Kerr black holes, asserts that the
quantum states in the near-horizon region of an extremal rotating
black hole are holographically dual to a two-dimensional chiral theory
whose Virasoro algebra arises as an asymptotic symmetry of the
near-horizon geometry.  In fact, in dimension $D$ there are $[(D-1)/2]$
commuting Virasoro algebras.  We consider a general canonical class of
near-horizon geometries in arbitrary dimension $D$, and show that in
any such metric the $[(D-1)/2]$ central charges each imply, via the
Cardy formula, a microscopic entropy that agrees with the
Bekenstein--Hawking entropy of the associated extremal black hole.  In
the remainder of the paper we show for most of the known rotating
black hole solutions of gauged supergravity, and for the ungauged
supergravity solutions with four charges in $D=4$ and three charges in
$D=5$, that their extremal near-horizon geometries indeed lie within
the canonical form.  This establishes that, in all these examples, the
microscopic entropies of the dual CFTs agree with the
Bekenstein--Hawking entropies of the extremal rotating black holes.

\vspace{15pt}

\thispagestyle{empty}

\pagebreak
\setcounter{page}{1}

\tableofcontents

\addtocontents{toc}{\protect\setcounter{tocdepth}{2}}

%%%%%%%%%%%%%%%%%%%%%%%%%%%%%%%%%%%%%%%%

\newpage
\section{Introduction}

   A recent paper \cite{guhasost} proposed a new holographic duality
symmetry in quantum gravity, in which the quantum states in the
near-horizon region of a four-dimensional extremal Kerr black hole are
identified with a certain two-dimensional chiral conformal field
theory.  This CFT arises by examining the asymptotic symmetry
generators associated with a class of diffeomorphisms of the
near-horizon Kerr geometry that obey suitably chosen boundary
conditions at infinity.  The Lie brackets of the infinitesimal
diffeomorphism transformations close on a centreless Virasoro algebra.
By defining charges associated with the transformations, and
evaluating the Dirac brackets of the charges, one obtains a Virasoro
algebra with a central charge that is related to the angular momentum
of the black hole.  By using the Cardy formula, the microscopic
entropy of the chiral CFT can be computed.  This calculation requires
that one invoke the ideas of Frolov and Thorne \cite{frotho} in order
to define a quantum theory in the extremal black hole geometry, and to
associate a non-zero temperature $T_{\rm FT}$ with the vacuum state.
It was shown in \cite{guhasost} that the microscopic entropy so
calculated agrees precisely with the Bekenstein--Hawking entropy of the
extremal Kerr black hole.  (See \cite{carlip,sol1,par1,par2,badbjesi,fagomosj1,fagomosj2} for some
earlier related work, and
\cite{hot,lumeipope,Azeyanagi:2008kb,hamunist,Nakayama:2008kg,kun1} for
recent follow-ups.)

   The proposed Kerr/CFT correspondence was extended to a wider class
of rotating black hole backgrounds in \cite{lumeipope}.  It was shown
that the microscopic entropy of the dual CFT again agrees with the
Bekenstein--Hawking entropy in the case of extremal Kerr--AdS black
holes, both in four dimensions and also in all higher dimensions.  A
new feature that arises in more than four dimensions is that there is
a Virasoro algebra, and a corresponding chiral CFT, associated with
each of the orthogonal 2-planes in which the black hole is rotating.
Curiously, although the central charges are different for the
different CFTs, their Frolov--Thorne temperatures differ too, in
precisely such a way that the Cardy formula leads to an identical
microscopic entropy for each of the CFTs.  Furthermore, each one of
these entropies agrees precisely with the Bekenstein--Hawking entropy
of the extremal rotating Kerr--AdS black hole \cite{lumeipope}.

   It is perhaps useful at this point to elaborate a little on the the
r\^ole of the Frolov--Thorne temperature in the calculation of
microscopic entropy via the Cardy formula.  The Cardy formula gives
the entropy of the two-dimensional CFT as
%%%%%
\be
S = 2\pi\, \sqrt{ \fft{c\, L}{6}}\,,\label{cardy}
\ee
%%%%%
where $c$ is the central charge and $L$ is the energy.  The temperature
of the CFT is then given by $dL=T dS$, and so from (\ref{cardy}) we have
$dS= \pi \sqrt{c/(6L)}\, T dS$ and hence 
%%%%%
\be
\sqrt{L} = \pi \sqrt{\fft{c}{6}}\, T\,.
\ee
%%%%
Substituting back into (\ref{cardy}) gives
%%%%%
\be
S = \fft{\pi^2}{3}\, c\, T\,.\label{cardy2}
\ee
%%%%%
It is in this form, with $c$ being the central charge of the Virasoro
algebra, and $T$ being the Frolov--Thorne expression for the
temperature of the near-horizon metric, that the Cardy formula
delivers an expression for the microscopic entropy of the CFT that
can be compared with the Bekenstein--Hawking entropy of the extremal
black hole.

   Another extension of the original proposal in \cite{guhasost} has
also recently been given, in which it was shown that the microscopic
entropy of the dual CFT agrees with the Bekenstein--Hawking entropy in
the case of the Kerr--Newman--(A)dS charged rotating extremal black hole
in four dimensions \cite{hamunist}.  It was also noted in
\cite{hamunist} that if one makes an assumption about the
Frolov--Thorne temperature for black hole solutions to a class of
four-dimensional theories involving the coupling of gravity to
electromagnetic and scalar fields, one could establish an equality of
the microscopic CFT entropy and the Bekenstein--Hawking entropy for a
wide class of higher-dimensional extremal black holes that are related
by dimensional reduction.

   In this paper, we shall probe the Kerr/CFT correspondence for a
large class of extremal higher-dimensional rotating charged black
holes.  Our strategy will be first to establish, for a general ansatz
for near-horizon geometries, a result that demonstrates the equality
of the microscopic entropy derived via the Cardy formula and the
Bekenstein--Hawking entropy.  Then, for any specific black hole
solution it only remains to construct its extremal near-horizon limit,
and to show that it is contained within the general ansatz mentioned
above, in order to establish the equality of the microscopic and the
Bekenstein--Hawking entropies for that case.

   The charged rotating black hole examples that we shall consider in
this paper include: the solution in four-dimensional $\mathcal{N} = 2$
(Einstein--Maxwell) gauged supergravity \cite{carter}; five-dimensional
minimal gauged supergravity \cite{cclp}; four-dimensional ungauged
supergravity with 4 unequal charges \cite{cvetyoum4}; four-dimensional
gauged supergravity with 2 sets of pairwise equal charges
\cite{chcvlupo2}; five-dimensional ungauged supergravity with 3
unequal charges \cite{cvetyoum}; five-dimensional gauged supergravity
with 3 charges, of which 2 are equal
\cite{chcvlupo3,chcvlupo4,meipope}; five-dimensional gauged
supergravity with both angular momenta equal and 3 charges
\cite{clpu1cube}; six-dimensional gauged supergravity \cite{chow6};
seven-dimensional gauged supergravity with two equal charges
\cite{chow7}; the higher-dimensional Kerr--AdS solution
\cite{gilupapo1, gilupapo2}; and a general class of black holes in
arbitrary dimension with two equal charges \cite{cvetyoum2,chow7}.

\section{General Extremal Rotating Black Holes and CFT Duals}

It was argued in \cite{hamunist} from the general structure of
four-dimensional extremal rotating black holes that the entropy of the
black hole can be obtained from the Cardy formula of the
two-dimensional conformal field theory in the boundary of the black
hole near-horizon geometry.  Here, we shall present a general argument
for higher-dimensional black holes.

   We consider first $D=5$ black holes that are
asymptotic to flat or AdS spacetimes, with the asymptotic
metric given by
%%%%
\be
ds^2=-(1 + \hat r^2\ell^{-2}) d\hat t^2 + 
\fft{d\hat r^2}{1 + \hat r^2\ell^{-2}} + \hat r^2
( d\theta^2 + \cos^2\theta d\hat\phi_1^2 + \sin^2\theta
d\hat \phi_2^2 )\,.
\ee
%%%
The discussion that follows is applicable for both vanishing and
non-vanishing cosmological constant $\ell^{-2}$.  In the extremal
limit, it is possible to extract the near-horizon geometry as an exact
solution in its own right, by first making the coordinate transformations
%%%%
\bea
&&\hat r = r_0(1 + \lambda\, \rho) , \qquad
\hat \phi_1= \phi_1 + \Omega^0_1\, \hat t\,,\qquad
\hat \phi_2= \phi_2 + \Omega^0_2\, \hat t\,,\qquad \hat t=
\fft{t}{2 \pi \dTH r_0 \lambda} .
\eea
%%%
Here $r_0$ is defined to be the horizon radius in the extremal limit.
The quantities $\Omega_i^0$ are the angular velocities on the horizon
for the two azimuthal angles $\hat \phi_i$, with the superscript 0
indicating that they are evaluated in the extremal limit.  Let $r_+$
be the outer horizon radius of the general non-extremal black hole, which we regard as one of the parameters of the general non-extremal family of solutions, and $T_H(r_+)$ be the corresponding Hawking temperature.  The quantity
$\dTH$ is defined to be \ben \dTH := \frac{\pd T_{\textrm{H}}}{\pd
r_+} \bigg| _{r_+ = r_0} .  \een For later purposes, we also define
%%%
\ben \dOi := \frac{\pd \Omega_i}{\pd r_+} \bigg| _{r_+ = r_0} ,
\een
%%%
where $\Omega_i(r_+)$ are the angular velocities for the general
non-extremal black hole.

     Taking the scaling parameter $\lambda$ to zero, we obtain the
near-horizon geometry of the extremal black hole, whose metric has the
form
%%%%
\bea
ds_5^2 &=& A(\theta) \left( -\rho^2 dt^2 + \fft{d\rho^2}{\rho^2} \right)
+ F(\theta) d\theta^2 + B_1(\theta)\, \tilde e_1^2 +
B_2 (\theta) (\tilde e_2 + C(\theta)\, \tilde e_1)^2\,,\nn\\
\tilde e_1 &=& d\phi_1 + k_1 \rho\, dt\,,\qquad
\tilde e_2 = d\phi_2 + k_2 \rho\, dt\,,\label{metform1}
\eea
%%%
where $A$, $B_i$, $C$ and $F$ are functions of the latitude coordinate
$\theta$.  The metric can be viewed as an $S^3$ bundle over AdS$_2$.
The AdS$_2$ base of the metric, written here in Poincar\'e coordinates
$(t,\rho)$, can be recast in global coordinates $(\tau,r)$ by means of
the transformations
%%%%%
\be
\rho=  r+ \sqrt{1+r^2}\, \cos\tau\,,\qquad
 t= \fft{\sqrt{1+r^2}\, \sin\tau}{r+ \sqrt{1+r^2}\, \cos\tau}\,,
\label{poinglob}
\ee
%%%%
Since this implies that $\rho dt= r d\tau + d\gamma$, where
%%%%%
\be
\gamma := \log \bigg( \fft{1+ \sqrt{1+r^2}\, \sin\tau}{\cos \tau +
r\, \sin\tau} \bigg) \,,\label{d2gamma}
\ee
%%%%%
it follows that if in addition we send $\phi_i\longrightarrow 
\phi_i-k_i\gamma$, then the metric (\ref{metform1}) becomes
%%%%
\bea
ds_5^2 &=& A(\theta) \left( -(1 + r ^2) dt^2 + \fft{dr^2}{1+r^2}
\right) + F(\theta) d\theta^2 + B_1(\theta)\, \tilde e_1^2 +
B_2 (\theta) (\tilde e_2 + C(\theta)\, \tilde e_1)^2\,,\nn\\
\tilde e_1 &=& d\phi_1 + k_1 r\, dt\,,\qquad
\tilde e_2 = d\phi_2 + k_2 r\, dt\,.\label{d5genh}
\eea
%%%%
In either form, the constants $k_1$ and $k_2$ are given by
%%%
\be
k_i=\fft{1}{2\pi T_i}\,,
\ee
%%%
with
%%%
\be
T_i = \lim_{r_+ \to r_0} \frac{T_{\textrm{H}}}{\Omega_i^0 - \Omega_i}
= - \frac{\dTH}{\dOi} . \label{Tidef}
\ee
%%%
The quantities $T_i$, defined first for higher-dimensional Kerr--AdS
black holes in \cite{lumeipope}, can be interpreted as the
Frolov--Thorne temperatures \cite{frotho,guhasost} associated with the
CFTs for each azimuthal angle $\phi_i$.  The Bekenstein--Hawking
entropy for the extremal black hole is given by
%%%
\be
S_{BH}= \ft14 \int d\theta \sqrt{B_1 B_2 F} \int d\phi_1
d\phi_2\,.
\ee
%%%%

     The five-dimensional near-horizon geometry (\ref{d5genh})
has a pair of commuting diffeomorphisms that generate two commuting 
Virasoro algebras:
%%%%
\bea
\zeta^\1_\n = - \ue^{-\im n \phi_1}\, \fft{\del}{\del\phi_1} - \im n\, r\,
   \ue^{-\im n \phi_1}\, \fft{\del}{\del r}\,,\nn\\
\zeta^\2_\n = - \ue^{-\im n \phi_2}\, \fft{\del}{\del\phi_2} - \im n\, r\,
   \ue^{-\im n \phi_2}\, \fft{\del}{\del r}\,.
\eea
%%%%%
The central charges $c_i$ in these Virasoro algebras, at the level of 
Dirac brackets
of the associated charges $Q^i_\n= 1/(8\pi)\int_{\del\Sigma} k^i_\n$,
can be calculated in the manner described in \cite{barbra, barcom} and applied in \cite{guhasost}, namely
from the $m^3$ terms in the expressions
%%%%%
\be
\fft1{8\pi}\, \int_{\del\Sigma} 
k_{\zeta^i_\m}[{\cal L}_{\zeta^i_\sst{(-m)}}g,g]=
 -\fft{\im}{12} (m^3 + \alpha m) c_i\,,\label{central}
\ee
%%%%%
where
%%%%%
\bea
k_\zeta[h,g] &=&\ft12 \Big[ \zeta_\nu\nabla_\mu h
  - \zeta_\nu \nabla_\sigma h_\mu{}^\sigma +
  \zeta_\sigma\nabla_\nu h_\mu{}^\sigma + \ft12 h \nabla_\nu\zeta_\mu
- h_\nu{}^\sigma \nabla_\sigma\zeta_\mu\nn\\
&&\qquad\qquad\qquad + \ft12 h_{\nu\sigma}
(\nabla_\mu\zeta^\sigma + \nabla_\sigma\zeta_\mu)\Big] \, {*(dx^\mu\wedge
dx^\nu)}\,,\label{kdef}
\eea
%%%%%
Taking $g_{\mu\nu}$ to be given by (\ref{d5genh}), we find that the
central charges are
%%%%
\be
c_i = \fft{3}{2\pi} k_i \int d\theta \sqrt{B_1 B_2 F} \int d\phi_1
d\phi_2 = \fft{6 k_i S_{BH}}{\pi}\,,
\ee
%%%
for $i=1$ and $i=2$.  Thus we have
%%%%
\be
S_{BH} = \fft{\pi^2}{3} c_1 T_1 = \fft{\pi^2}{3} c_2 T_2\,,
\ee
%%%%
in precise agreement with the microscopic entropy 
given by the Cardy formula (\ref{cardy2}).

   The argument above can be straightforwardly generalised to higher
dimensions.  The near-horizon geometry of extremal rotating black
holes in $D=2n+\epsilon$ dimensions, with $\epsilon=0,1$, can be
written, using Poincar\'{e} AdS$_2$ coordinates, as
%%%%
\bea
ds^2 &=& A \left( - \rho^2 \, \ud t^2 + \frac{\ud \rho^2}{\rho^2} \right) +
\sum_{\alpha=1}^{n-1} F_\alpha \, \ud y_\alpha^2 + \sum_{i,j = 1}^{n-1+\epsilon}
\tilde g_{ij}\, \tilde e_i \, \tilde e_j\,,\nn\\
\tilde e_i & = & \ud \phi_i + k_i \rho \, \ud t \,,\qquad
k_i=\fft{1}{2\pi T_i}\,,\qquad
T_i= - \frac{\dTH}{\dOi} ,\label{dgenh}
\eea
%%%
or alternatively, using global AdS$_2$ coordinates, as
\bea
ds^2 & = & A \left( -(1 + r^2) \, \ud \tau^2 + \fft{\ud r^2}{1 + r^2} \right) +
\sum_{\alpha=1}^{n-1} F_\alpha \, \ud y_\alpha^2 + \sum_{i,j = 1}^{n-1+\epsilon}
\tilde g_{ij}\, \tilde e_i \, \tilde e_j \,,\nn\\
\tilde e_i & = & \ud \phi_i + k_i r \, \ud \tau\,,\qquad
k_i=\fft{1}{2\pi T_i}\,,\qquad
T_i= - \frac{\dTH}{\dOi} .
\eea
%%%
Here we follow \cite{chlupo} and use a set of unconstrained
latitudinal coordinates $y_\alpha$, rather than the direction cosines
$\mu_a$ subject to $\sum_{a = 1}^n \mu_a^2 = 1$ that were used in the
original formulation of the higher-dimensional Ricci-flat
\cite{myeper} or asymptotically AdS \cite{gilupapo1,gilupapo2}
rotating black holes.  The functions $A$, $F_\alpha$ and
$\tilde{g}_{ij}$ depend only on these latitudinal coordinates.  The
metric has $n-1+\epsilon$ copies of the Virasoro algebra.  It has been shown that near-horizon geometries are generally of this form for classes of theories that are of interest in four and five dimensions \cite{kulure}, and also for cohomogeneity-1 horizons in arbitrary dimension \cite{kun3}.  We have
verified for dimensions $D \leq 7$ that the central charges are given
by
%%%%
\bea
c_i &=& \fft{3}{2\pi} k_i \int d^{n-1} y_\alpha
\bigg( \det{\tilde g_{ij}}\prod_{\alpha = 1}^{n-1} F_\alpha \bigg) ^{1/2}
\int d\phi_1\cdots d\phi_{n-1+\epsilon}\nn\\
&=& \fft{6k_i S_{BH}}{\pi}\,.
\eea
Since this relation does not have any features relying on a particular
dimension, it is very likely to hold in arbitrary dimension.  It
follows that
%%%
\be
S_{BH}= \ft13 \pi^2 c_i T_i\,,\qquad\hbox{for each $i$},
\label{cardygen}
\ee
%%%
holds in general, in complete agreement with the microscopic entropy
given by the Cardy formula (\ref{cardy2}).

     In the next few sections, we shall examine a large class of
charged rotating black holes in diverse dimensions.  We obtain the
near-horizon geometries of these black holes in the extremal
limit.  We demonstrate that the metrics can all be cast
into the form (\ref{dgenh}), and hence that the Cardy formulae are
all satisfied.

\section{Einstein--Maxwell AdS Supergravities in Four
and Five Dimensions}

We shall start our main discussion with two relatively simple
examples, namely the charged rotating black holes
in Einstein--Maxwell AdS supergravities in four and five
dimensions.

\subsection{Four-dimensional Einstein--Maxwell AdS supergravity}

This example, the Kerr--Newman--AdS solution, was discussed in detail
in \cite{hamunist}; we include it here for completeness.  The metric
is given by
%%%%
\bea
ds^2&=& \rho^2 \left( \fft{d\hat r ^2}{\Delta} + \fft{d\theta ^2}{\Delta_{\theta}}
\right) + \fft{\Delta_\theta \sin^2\theta}{\rho^2} \left( a d\hat t -
\fft{\hat r^2 + a^2}{\Xi} d\hat \phi \right) ^2 -
\fft{\Delta}{\rho^2} \left( d\hat t - \fft{a \sin^2\theta}{\Xi} d\hat \phi
\right) ^2\,,\nn\\
\rho^2&=& \hat r^2 + a^2 \cos^2\theta\,,\qquad
\Delta=(\hat r^2 + a^2) (1 + \hat r^2\ell^{-2}) - 2 M \hat r + Q^2\,,\nn\\
&&
\Delta_\theta = 1 - a^2\ell^{-2} \cos^2\theta\,,\qquad
\Xi=1 - a^2\ell^2\,.
\eea
%%%
Here $Q^2=p^2 + q^2$, with $(q,p)$ being the electric and magnetic
charges.  The solution describes a charged black hole with the outer
horizon at $\hat r=r_+$, where $r_+$ is the largest root of the
function $\Delta (\hat{r})$.  The metric is asymptotically AdS$_4$ in
global coordinates, but with non-vanishing angular velocity
$\Omega_\infty=-a^2\ell^{-2}$.  The Hawking temperature, entropy and
angular velocity on the horizon are given by
%%%%
\bea
T_H&=&\fft{r_+^2 -a^2 - Q^2 + r_+^2 \ell^{-2}
(3r_+^2 + a^2)}{4\pi r_+ (r_+^2 + a^2)}\,,\qquad
\Omega_\phi=\fft{\Xi a}{r_+^2 + a^2}\,,\nn\\
S&=&\fft{\pi (r_+^2 + a^2)}{\Xi}\,.
\eea
%%%

      The extremal limit is achieved when the parameters $M$ and $Q$ take
the following values:
%%%%
\be
M=r_0 + r_0 (2r_0^2 + a^2) \ell^{-2}\,,\qquad
Q^2=r_0^2 - a^2 + r_0^2 (3r_0^2 + a^2) \ell^{-2}\,.
\ee
%%%
The horizon of the metric is at $\hat{r} = r_0$, with the function
$\Delta$ near the horizon given by
%%%
\be
\Delta = V (\hat r-r_0)^2 + \textrm{O} (\hat r-r_0)^3 \,,\qquad
\hbox{with}\qquad
V=1 + (6 r_0^2 + a^2)\ell^{-2}\,.
\ee
%%%%
To obtain the near-horizon geometry, we make the coordinate
transformation
%%%
\be
\hat r = r_0(1 + \lambda \rho)\,,\qquad
\hat\phi = \phi + \Omega_\phi^0 \hat t\,,
\ee
%%%
where $\Omega_\phi^0=\Omega_\phi|_{r_+=r_0}$.  We then scale the
time coordinate $\hat t$ by
%%%
\be
\hat t = \fft{r_0^2 + a^2}{r_0V \lambda}t\,,
\ee
%%%
and send $\lambda\rightarrow 0$.  We obtain the metric
%%%%
\bea
ds^2 &=& \fft{\rho_0^2}{V} \left( -\rho^2 dt^2 +
\fft{d\rho^2}{\rho^2} + \fft{Vd\theta^2}{\Delta_\theta} \right)
+\fft{(r_0^2 + a^2)^2 \sin^2\theta\, \Delta_\theta}{\Xi^2\rho_0^2}
\left( d\phi + \fft{1}{2\pi T_\phi} \rho\, dt \right) ^2\,,\nn\\
\rho_0^2&=&r_0^2 + a^2\cos^2\theta\, ,
\eea
%%%
where the Frolov--Thorne temperature $T_\phi$ is given by
%%%%
\bea
T_\phi = -\fft{\del_{r_+} T_H}{\del_{r_+} \Omega_\phi} \bigg|_{r_+=r_0}
=\fft{V (r_0^2 + a^2)}{4 \pi \Xi a r_0}\,.
\eea
%%%
The entropy in the extremal limit is
%%%
\be
S=\fft{\pi (r_0^2 + a^2)}{\Xi}\,.
\ee
%%%
The central charge can be easily obtained, given by
%%%%
\be
c=\fft{12a r_0}{V}\,.
\ee
%%%

\subsection{Five-dimensional minimal gauged supergravity}

The general non-extremal rotating black hole in five-dimensional
minimal gauged supergravity with two arbitrary angular momenta was
obtained in \cite{cclp}.  Here we shall adopt the notation given in
\cite{chowsym}.  The metric is given by
%%%%
\be
ds^2 = - e^0 e^0 + \sum_{i = 1}^4 e^i e^i ,
\ee
%%%
where
%%%
\bea
e^0 & = & \sqrt{\frac{R}{\hat r^2 + y^2}} {\mathcal{A}} , \quad e^1 =
\sqrt{\frac{\hat r^2 + y^2}{R}} d r , \quad e^2 =
\sqrt{\frac{Y}{\hat r^2 + y^2}} (dt' - \hat r^2 d \psi_1) ,
\\
e^3 & = & \sqrt{\frac{\hat r^2 + y^2}{Y}} dy , \quad e^4 =
\frac{a b}{\hat r y} \left( d t' + (y^2 - \hat r^2) d\psi_1 -
 \hat r^2 y^2 d\psi_2 + \frac{q y^2}{a b (\hat r^2 + y^2)} {\mathcal{A}}
\right ) ,\nn
\eea
%%%
and
%%%
\bea
&& R = \frac{(1 + \hat r^2\ell^{-2}) (\hat r^2 + a^2) (\hat r^2 + b^2) +
2abq + q^2}{\hat r^2} - 2M\,, \\\
&&Y = - \frac{(1 - y^2\ell^{-2})
(a^2 - y^2) (b^2 - y^2)}{y^2} ,\qquad {\mathcal{A}} = dt' +
y^2 d\psi_1 .\nn
\eea
%%%%
The coordinates $t'$, $\psi_1$ and $\psi_2$ are not proper canonical
time and azimuthal coordinates.  The proper coordinates $(\hat t, \hat
\phi_1, \hat \phi_2)$ are given by
%%%%
\bea
t'&=&\hat t -(a^2 + b^2) \psi_1 - a^2 b^2 \psi_2\,,\nn\\
\psi_1 &=& \fft{a \hat \phi_1}{\Xi_a(a^2-b^2)} +
\fft{b \hat \phi_2}{\Xi_b (b^2-a^2)}\,,\nn\\
\psi_2&=& \fft{\hat \phi_1}{a(b^2-a^2)\Xi_a}+
\fft{\hat \phi_2}{b(a^2-b^2)\Xi_b}\,,
\eea
%%%%
where $\Xi_a = 1 - a^2 \ell^{-2}$ and $\Xi_b = 1 - b^2 \ell^{-2}$.
Then the coordinates $\phi_1$ and $\phi_2$ have period $2\pi$.  The
metric is AdS$_5$ asymptotically, but in a rotating coordinate frame
with angular velocities $\Omega_1=-a\ell^{-2}$ and
$\Omega_2=-b\ell^{-2}$.  The thermodynamic quantities for this black
hole were obtained in \cite{cclp}.  Here we shall present the
temperature, entropy and the angular velocities of the horizon.  These
are given by
%%%%
\bea
T_H&=&\fft{r_+^2}{4\pi [r_+^4 + (a^2 + b^2) r_+^2 + a b (a b + q) ]}
\left( \fft{\del R}{\del \hat r} \right) \bigg|_{\hat r = r_+}\,,\nn\\
S&=& \fft{\pi^2 [r_+^4 + (a^2 + b^2) r_+^2 + a b (a b + q)]}{
2r_+\Xi_a\Xi_b}\,,\nn\\
\Omega_1 &=& \fft{\Xi_a (a r_+^2 + a b^2 + q b)}{(r_+^2 + a^2)(r_+^2 + b^2)
+ q a b}\,,\qquad
\Omega_2= \fft{\Xi_b (b r_+^2 + a^2 b + q a)}{(r_+^2 + a^2)(r_+^2 + b^2)
+ q a b} .
\eea
%%%%
We now consider the extremal limit, given by the following conditions:
%%%%
\bea
M &=&\fft{(1 + r_0^2 \ell^{-2}) (r_0^2 + a^2) (r_0^2 + b^2) +
q^2 + 2abq}{2r_0^2}\,,\nn\\
\ell^{-2} &=& \fft{(a b + q)^2 - r_0^4}{r_0^4 (a^2 + b^2 + 2r_0^2)}\,.
\eea
%%%%
Near the horizon, we have
%%%
\be
R = V (\hat r-r_0)^2 + \textrm{O}(\hat{r} - r_0)^3 \,,
\qquad V= \ft12 R''(r_0)\,.
\ee
%%%
To extract the near-horizon geometry, we make the following coordinate
transformation:
%%%
\be
\hat r = r_0(1 + \lambda \rho) \,,\qquad
\hat\phi_1=\phi_1 + \Omega_1^0\, \hat t\,,\qquad
\hat \phi_2=\phi_2 + \Omega_2^0\, \hat t\,,
\ee
%%%
where $\Omega_i^0=\Omega_i|_{r_+=r_0}$.
We then scale the time coordinate $\hat t$ as
%%%
\be
\hat t = \beta t\,,\qquad
\beta = \fft{1}{2 \pi r_0 \dTH \lambda} =
\fft{r_0^4 + (a^2 + b^2) r_0^2 + a b (a b + q)}{V r_0^3 \lambda}
\,.
\ee
%%%%
Taking the limit of $\lambda\rightarrow 0$, the vielbeins become
%%%
\bea
e^0 &=&\sqrt{\frac{r_0^2 + y^2}{V}} \rho dt, \quad e^1 =
\sqrt{\frac{r_0^2 + y^2}{V}} \fft{d\rho}{\rho} ,
\quad e^3 = \sqrt{\frac{r_0^2 + y^2}{Y}} dy ,\nn\\
e^2 &=&
\sqrt{\frac{Y}{r_0^2 + y^2}} \left( \fft{a (a^2 + r_0^2)
\,\tilde e_1}{\Xi_a (a^2-b^2)} +
\fft{b(b^2 + r_0^2)\, \tilde e_2}{\Xi_b (b^2-a^2)} \right) ,
\nn\\
e^4 &=&
\frac{a b}{r_0 y} \bigg(
\fft{(a^2-y^2) (aq y^2 +
b (a^2 + r_0^2)(r_0^2 + y^2))}{ab (a^2-b^2) \Xi_a (r_0^2 + y^2)}
\,\tilde e_1\nn\\
&& \qquad\qquad+
\fft{(b^2-y^2) (bq y^2 + a (b^2 + r_0^2)(r_0^2 + y^2))}{ab (b^2-a^2)
\Xi_b (r_0^2 + y^2)}\, \tilde e_2 \bigg) ,
\eea
%%%%
where
%%%
\be
\tilde e_{i} = d\phi_i + k_i \rho dt\,,\qquad
k_i=\fft{1}{2\pi T_i}\,,
\ee
%%%
and $T_i$'s are the Frolov--Thorne temperatures defined in (\ref{Tidef}).
Thus we see that the near-horizon geometry of the extremal black hole
can be put in the general form (\ref{dgenh}) discussed in section 2,
and hence the
Cardy formulae (\ref{cardygen}) are satisfied.  To be specific, we have
%%%%
\be
T_1 = \fft{r_0 V [ (r_0^2 + a^2)(r_0^2 + b^2) + q a b ]}{4\pi
\Xi_a [ a(r_0^2 + b^2)^2 + q b (b^2 + 2 r_0^2) ]}
\,,\quad
T_2 = \fft{r_0 V [ (r_0^2 + a^2)(r_0^2 + b^2) + q a b ]}{4\pi
\Xi_b [ b(r_0^2 + a^2)^2 + q a (a^2 + 2 r_0^2) ]}
\,.
\ee
%%%
The corresponding central charges are given by
%%%%
\be
c_1 = \fft{6\pi [a (r_0^2 + b^2)^2 + q b (b^2 + 2 r_0^2)]}{r_0^2
\Xi_b V}\,,\quad
c_2 = \fft{6\pi [b (r_0^2 + a^2)^2 + q a (a^2 + 2 r_0^2)]}{r_0^2
\Xi_a V}\,.
\ee
%%%

\section{Four Dimensions}

    In this and following sections, we consider a variety of rotating
black holes involving multiple charges in various dimensions.  We
start here with four dimensions, and then later proceed to increase
the dimensionality.

\subsection{Ungauged supergravity with four unequal charges}

  Black holes with four unequal charges arise from the bosonic sector
of the four-dimensional ${\cal N}=2$ ungauged supergravity coupled to
three vector multiplets.  The metric was first obtained in
\cite{cvetyoum4}, and the explicit form of the gauge potentials was
given in \cite{chcvlupo2}.

  The solution is specified by mass, angular momentum, and
two electric and two magnetic charges.  The metric takes the form
\be
ds_4^2 = -\fft{\rho^2-2m\hat{r}}{W}\, (d{\hat t}+ B\, d{\hat \phi})^2 +
    W\, \bigg( \fft{d \hat{r}^2}{\Delta} + d\theta^2 +
   \fft{\Delta\, \sin^2\theta\, d{\hat \phi}^2}{\rho^2-2m\hat{r}} \bigg) \, ,
\label{4dmetric1}
\ee
%%%%%
where
\bea
\Delta &=& \hat{r}^2 -2 m \hat{r} + a^2\,,\qquad \rho^2=
\hat{r}^2+a^2\cos^2\theta\, , \nn\\
B &=& \fft{2m(a^2-u^2)[\hat{r}c_{1234}-
                      (\hat{r}-2m)s_{1234}]}{a(\rho^2 -2m\hat{r})}\,,\nn\\
W^2&=&r_1\, r_2\, r_3\, r_4 + u^4 + u^2 [2 \hat{r}^2 + 2 m
\hat{r} (s_1^2 + s_2^2
+ s_3^2 + s_4^2)\nn\\
&& + 8m^4 c_{1234}s_{1234} -
4m^4(s_{123}^2 + s_{124}^2 + s_{134}^2 +s_{234}^2 +2
s_{1234}^2)]\,,\nn\\
r_i &=& \hat{r} + 2 m s_i^2\,,\qquad u=a\cos\theta\,,\nn\\
c_{i_1\cdots i_n}&=& \cosh\delta_{i_1}\cdots\cosh\delta_{i_n}\,,\qquad
s_{i_1\cdots i_n} = \sinh\delta_{i_1}\cdots\sinh\delta_{i_n}\,.
\label{variousdefs}
\eea
%%%%
The outer and inner horizons are at $\hat{r} = r_\pm$, with
\be
r_{\pm}=m\pm \sqrt{m^2-a^2}\, .
\ee
The entropy $S$,  Hawking temperature $T_H$ and the angular velocity
$\Omega$ have the explicit form
\bea
S&=&  {2\pi}~
[ m^2 (c_{1234} + s_{1234})
+ m\sqrt{m^2-a^2}
(c_{1234} - s_{1234})]~\,,\nn\\
T_H & = & \frac{1}{4 \pi m [c_{1234} - s_{1234}
+ (c_{1234} + s_{1234}) m / \sqrt{m^2 - a^2}]}~\,,\nn\\
\Omega & = &2\pi T_H \frac{a}{\sqrt{m^2-a^2}}~.
\eea
The extreme black hole corresponds to
\be m=a\, , \ \ \  {\hbox{and}} \ \ r_+=a\, .
\ee
The near-horizon geometry of the extreme black hole is obtained by taking
\be
\hat{r} = a (1 + \lambda \rho) \,, \ \ \ \hat{\phi} = \phi + \Omega \hat{t} , \quad \hat{t} = \frac{t}{\lambda} ,
\ee
with $\lambda \to 0$.  The near-horizon metric is then
\be
ds_4^2 = W_0\, \left( -\rho^2 dt^2 + \fft{d\rho^2}{\rho^2} +
d\theta^2 \right) \, +\frac{a^2 \sin^2 \theta \, B_0^2}{W_0}\,
(d\phi + k \rho \, dt)^2\,. \label{4dmetric2}
\ee
where
\bea
B_0&=&B | _{\hat{r} = a, m = a}=-2a(c_{1234} - s_{1234}) \, , \nn\\
k&=&\fft{1}{2\pi T_\phi}=
- \frac{\partial_{r_+} \Omega}{2\pi\partial_{r_+} T_H}\bigg|_{r_+=a}=
\frac{c_{1234} - s_{1234}}{c_{1234} + s_{1234}}\, , \nn\\
 W_0&=& W|_{\hat{r} = a, m = a}\, .
 \eea
%%%
Thus, we see that the form of the near-horizon geometry of the
extremal black hole fits into the general pattern discussed in section
2, and hence the Cardy formula is satisfied.

\subsection{U(1)$^4$ gauged supergravity with pairwise equal charges}

The most general charged rotating black hole solution known in
four-dimensional U(1)$^4$ gauged supergravity has the four U(1)
charges pairwise equal \cite{chcvlupo2}.

The metric is
\bea
\ud s^2 & = & H \bigg[ - \frac{R}{H^2 (\hat{r}^2 + y^2)}
\left( \ud \hat{t} - \frac{a^2 - y^2}{\Xi a} \ud \hat{\phi} \right) ^2 +
\frac{\hat{r}^2 + y^2}{R} \ud \hat{r}^2 + \frac{\hat{r}^2 + y^2}{Y}
\ud y^2 \nnr
&& + \frac{Y}{H^2 (\hat{r}^2 + y^2)} \left( \ud \hat{t} -
\frac{(\hat{r} + q_1) (\hat{r} + q_2) + a^2}{\Xi a} \ud \hat{\phi}
\right) ^2 \bigg] ,
\eea
where
\bea
&& R = \hat{r}^2 + a^2 + g^2 (\hat{r} + q_1) (\hat{r} + q_2)
[(\hat{r} + q_1) (\hat{r} + q_2) + a^2] - 2 m \hat{r} , \nnr
&& Y = (1 - g^2 y^2) (a^2 - y^2) , \quad \Xi = 1 - a^2 g^2 , \nnr
&& H = \frac{(\hat{r} + q_1) (\hat{r} + q_2) +
y^2}{\hat{r}^2 + y^2} , \quad q_I = 2 m s_I^2 ,
\quad s_I = \sinh \delta_I .
\eea
%%%
Note that, as is standard in the gauged supergravity literature, we
are using $g$ to denote the gauge-coupling constant, which is related
to the AdS length scale $\ell$ by $g = \ell^{-1}$.  We have used a
shifted azimuthal coordinate $\hat{\phi}$ that gives an asymptotically
rotating coordinate frame; the coordinate change $\hat{\phi}
\rightarrow \hat{\phi} - a g^2 \hat{t}$ would give an asymptotically
non-rotating coordinate frame.  This shifted azimuthal coordinate is
used merely to make the metrics more convenient to write, and is not
otherwise significant.  The Hawking temperature and entropy are
%%%%
\bea
T_{\textrm{H}} & = & \frac{R'|_{\hat{r} = r_+}}{4 \pi
[(r_+ + q_1) (r_+ + q_2) + a^2]} \nnr
& = & \frac{r_+^2 - a^2 + a^2 g^2 (r_+^2 - q_1 q_2) +
g^2 (r_+ + q_1) (r_+ + q_2) (3 r_+^2 + q_1 r_+ + q_2 r_+ -
q_1 q_2)}{4 \pi r_+ [(r_+ + q_1) (r_+ + q_2) + a^2]} , \nnr
S & = & \frac{\pi [(r_+ + q_1) (r_+ + q_2) + a^2]}{\Xi} .
\eea
In our asymptotically rotating coordinate frame, the angular velocity
of the horizon is
%%%%
 \ben
\hat{\Omega} = \frac{\Xi a}{(r_+ + q_1) (r_+ + q_2) + a^2} .
\een

For an extremal solution, with a horizon at $\hat{r} = r_0$, we have
$R|_{\hat{r} = r_0} = 0$ and $R'|_{\hat{r} = r_0} = 0$, and so
%%%
\ben
r_0^2 - a^2 + a^2 g^2 (r_0^2 - q_1 q_2) +
g^2 (r_0 + q_1) (r_0 + q_2) (3 r_0^2 + q_1 r_0 + q_2 r_0 - q_1 q_2) = 0 .
\een
Then we have the near-horizon expansion
\ben
R = V (\hat{r} - r_0)^2 + \textrm{O} (\hat{r} - r_0)^3 ,
\een
where
%%%
\ben V = 1 + g^2 (6 r_0^2 + 6 q_1 r_0 + 6 q_2 r_0 + a^2 + q_1^2 +
q_2^2 + 4 q_1 q_2) .  \een
%%%
To obtain the near-horizon geometry, we make the coordinate changes
\ben
\hat{r} = r_0(1 + \lambda \rho)\,, \qquad \hat{\phi} = \phi +
\Omega^0 \hat{t} , \qquad \hat{t} = \frac{t}{2 \pi \dTH r_0 \lambda} ,
\een
and then take the limit $\lambda \rightarrow 0$.  The near-horizon
geometry is
\bea
\ud s^2 & = & H_0 \bigg[ (r_0^2 + y^2) \left( - \rho^2 \ud t^2 +
\frac{\ud \rho^2}{\rho^2} \right) + \frac{r_0^2 + y^2}{Y} \ud y^2 \nnr
&& + \frac{Y}{H_0^2 (r_0^2 + y^2)} \left( \frac{2 r_0 + q_1 + q_2}{V}
\rho \, \ud t + \frac{(r_0 + q_1) (r_0 + q_2) + a^2}{\Xi a}
\ud \phi \right) ^2 \bigg] ,
\eea
%%%
where $H_0 = H | _{\hat{r} = r_0}$.  This can be cast in the form of (\ref{dgenh}), so the Cardy formulae are satisfied.

For the extremal solution, the Frolov--Thorne temperatures are
%%%%
\ben
T_0 = 0 , \qquad T_1 = \frac{V [(r_0 + q_1) 
(r_0 + q_2) + a^2]}{2 \pi \Xi a (2 r_0 + q_1 + q_2)} .
\een
The central charge is 
\ben
c_1 = \frac{6 a (2 r_0 + q_1 + q_2)}{V} .
\een

\section{Five Dimensions}

\subsection{Ungauged supergravity with three unequal charges}

The $U(1)^3$ charged black hole in the STU model in $D=5$ ungauged
supergravity was obtained in \cite{cvetyoum}.  The solution
was expressed in a simpler form in \cite{chcvlupo4},
in which the metric is given by
%%%%
\bea
ds_5^2 &=& (H_1 H_2 H_3)^{1/3} (x+y) \left( - \fft{G}{(x+y)^3 H_1 H_2 H_3}
(d\hat t + {\cal A})^2 + ds_4^2 \right) \,,\nn\\
ds_4^2 &=& \left( \fft{dx^2}{4 X} + \fft{dy^2}{4Y} \right)
    + \fft{U}{G}\, \left( d\chi - \fft{Z}{U}\, d\sigma \right) ^2 +
     \fft{X Y}{U}\, d\sigma^2\,,\label{d5cymet}
\eea
%%%%
where
%%%%%
\bea
X &=& (x+a^2)(x+b^2)- 2Mx \,,\qquad Y= - (a^2-y)(b^2-y) \,,\nn\\
G &=& (x+y)(x+y-2M)  \,,\qquad
U = y X - x Y \,,\qquad Z = a b (X+Y) \,,\nn\\
{\cal A}&=& \fft{2M c_1 c_2 c_3}{x+y-2M} [(a^2+b^2-y) d\sigma -a b d\chi]
- \fft{2M s_1 s_2 s_3}{x+y}\, (ab d\sigma - y d\chi)\,,\nn\\
H_i &=& 1 + \fft{2M s_i^2}{x+y}\,,\qquad
s_i=\sinh \delta_i\,,\qquad
c_i=\cosh \delta_i\,.
\eea
%%%%%
Here, $x$ is the radial coordinate with the asymptotic flat region at
$x=\infty$.  The horizon is at $x=x_+$, where $x_+$ is the largest
root of $X$.  The latitude coordinate $y$ runs from $a$ to $b$.
The $U(1)$ coordinates $\sigma$ and $\chi$ are related to the canonical
azimuthal coordinates as follows:
%%%%
\be
\sigma = \fft{a\, \hat \phi_1 - b\, \hat \phi_2}{a^2-b^2}\,,\qquad
\chi=\fft{b\, \hat \phi_1 - a\, \hat \phi_2}{a^2-b^2}\,.
\ee
%%%%
The entropy, Hawking temperature, and angular velocities on the horizon
are given by
%%%
\bea
S&=& \fft{\pi^2 (x_+ + a^2)(x_+ + b^2) (c_1 c_2 c_3\, x_+ +
s_1 s_2 s_3\, ab)}{2 x_+^{3/2}}\,,\nn\\
T_H&=& \fft{\sqrt{x_+} (x_+^2 - a^2 b^2)}{2\pi (x_++a^2)(x_++b^2)
(c_1 c_2 c_3\, x_+ + s_1 s_2 s_3\, ab)}\,,\\
\Omega_1&=&\fft{a x_+}{(x_+ + a^2)(c_1 c_2 c_3\, x_+ + s_1 s_2 s_3\,a b)}
\,,\quad
\Omega_2=\fft{b x_+}{(x_+ + b^2)(c_1 c_2 c_3\, x_+ + s_1 s_2 s_3\,ab)}
\,.\nn
\eea
%%%

     The extremal limit of the solution is achieved with the
condition $M=\ft12 (a + b)^2$, in which case the horizon is
at $x=x_0$, where $x_0=a b$.  As in the previous case, the
extremal limit can be extracted by the following coordinate
transformation:
%%%
\be
x = x_0(1 + \lambda\, \rho)\,,\qquad \hat \phi_i = \phi +
\Omega_i^0\, \hat t\,,\qquad
\hat t = \fft{t}{2\pi x_0 T'_H(x_0) \lambda}\,.
\ee
%%%
We then take the limit $\lambda\rightarrow 0$.  The near-horizon
geometry then has the following form:
%%%%
\be
ds_5^2 = \frac{ab + y}{4} \prod_{i=1}^3
\left( 1 + \fft{(a+b)^2s_i^2}{ab+y} \right) ^{1/3}
\bigg( -\rho^2 dt^2 + \fft{d\rho^2}{\rho^2} + \fft{dy^2}{(a^2-y)(b^2-y)}
+ \sum_{i,j = 1}^2 \tilde g_{ij}(y) \tilde e_i\tilde e_j \bigg) \,,
\ee
where $\tilde{e}_i = d \phi_i + k_i \rho\, dt$, with $k_i=1/(2\pi
T_i)$.  This is precisely the same form as in (\ref{dgenh}), and so
the Cardy formulae are satisfied.  Here we present the entropy, the
Frolov--Thorne temperatures and central charges:
%%%
\bea
S &=& \ft12 \pi^2 (a+b)^2 \sqrt{ab} (c_1c_2c_3 + s_1 s_2 s_3)\,,\nn\\
T_1 &=& \fft{\sqrt{ab} (c_1c_2c_3 + s_1s_2s_3)}{\pi (b c_1 c_2 c_3
-a s_1 s_2 s_3)}\,,\qquad
T_2 = \fft{\sqrt{ab} (c_1c_2c_3 + s_1s_2s_3)}{\pi (a c_1 c_2 c_3
-b s_1 s_2 s_3)}\,,\nn\\
c_{\phi_1} &=& \ft32\pi (a+b)^2 (bc_1 c_2 c_3 - a s_1 s_2 s_3)\,,\quad
c_{\phi_2} = \ft32\pi (a+b)^2 (ac_1 c_2 c_3 - b s_1 s_2 s_3)\,.
\eea
%%%

\subsection{U(1)$^3$ gauged supergravity}

The maximal five-dimensional gauged supergravity has gauge group
SO$(6)$, which has Cartan subgroup U(1)$^3$.  We have already
considered black holes in minimal gauged supergravity, which
corresponds to all three abelian charges being equal.  Here, we
consider some further black hole solutions of the theory.

\subsubsection{Charge parameters
$\delta_1 = \delta_2$, $\delta_3 = 0$}

Another particularly simple charged and rotating black hole in
five-dimensional U(1)$^3$ gauged supergravity has three charge
parameters $\delta_I$ that satisfy $\delta_1 = \delta_2 =: \delta$ and
$\delta_3 = 0$, as well as both angular momenta independent \cite{chcvlupo3}.

The metric, using the vielbeins presented in \cite{chow7} but here with Boyer--Lindquist azimuthal coordinates, is
%%%
\bea
\ud s^2 & = & H^{2/3} \bigg[ - \frac{R}{H^2 (\hat{r}^2 + y^2)}
\mathcal{A} ^2 + \frac{\hat{r}^2 + y^2}{R} \ud \hat{r}^2 +
\frac{\hat{r}^2 + y^2}{Y} \ud y^2 \nnr
&& + \frac{Y}{\hat{r}^2 + y^2} \left( \ud \hat{t} -
\frac{a (\hat{r}^2 + a^2)}{\Xi_a (a^2 - b^2)} \ud \hat{\phi}_1 -
\frac{b (\hat{r}^2 + b^2)}{\Xi_b (b^2 - a^2)} \ud \hat{\phi}_2 -
\frac{q}{H (\hat{r}^2 + y^2)} \mathcal{A} \right) ^2 \nnr
&& + \frac{a^2 b^2}{\hat{r}^2 y^2} \left( \ud \hat{t} -
\frac{(\hat{r}^2 + a^2) (a^2 - y^2)}{\Xi_a a (a^2 - b^2)}
\ud \hat{\phi}_1 - \frac{(\hat{r}^2 + b^2)
(b^2 - y^2)}{\Xi_b b (b^2 - a^2)}
\ud \hat{\phi}_2 - \frac{q}{H (\hat{r}^2 + y^2)} \mathcal{A} \right) ^2 \bigg] , \nnr
\eea
where
%%%
\bea && R = \frac{(\hat{r}^2 + a^2) (\hat{r}^2 + b^2)}{\hat{r}^2} +
g^2 (\hat{r}^2 + a^2 + q) (\hat{r}^2 + b^2 + q) - 2 m , \nnr && Y = -
\frac{(1 - g^2 y^2) (a^2 - y^2) (b^2 - y^2)}{y^2} , \quad \Xi_a = 1 -
a^2 g^2 , \quad \Xi_b = 1 - b^2 g^2 , \nnr && H = 1 +
\frac{q}{\hat{r}^2 + y^2} , \quad q = 2 m s^2 , \quad s = \sinh \delta
, \nnr && \mathcal{A} = \ud \hat{t} - \frac{a (a^2 - y^2)}{\Xi_a (a^2
- b^2)} \ud \hat{\phi}_1 - \frac{b (b^2 - y^2)}{\Xi_b (b^2 - a^2)} \ud
\hat{\phi}_2 .  
\eea 
%%%
We have used shifted azimuthal coordinates
$\hat{\phi}_i$ that give an asymptotically rotating coordinate frame;
the coordinate changes $\hat{\phi}_1 \rightarrow \hat{\phi}_1 - a g^2
\hat{t}$ and $\hat{\phi}_2 \rightarrow \hat{\phi}_2 - b g^2 \hat{t}$
would give an asymptotically non-rotating coordinate frame.  The
Hawking temperature and entropy are
%%%%
\bea
T_{\textrm{H}} & = & \frac{(\hat{r}^2
R)'|_{\hat{r} = r_+}}{4 \pi [(r_+^2 + a^2) (r_+^2 + b^2) + q r_+^2]} =
\frac{r_+^4
- a^2 b^2 + g^2 r_+^4 (2 r_+^2 + a^2 + b^2 + 2 q)}{2 \pi r_+ [(r_+^2 +
a^2) (r_+^2 + b^2) + q r_+^2]} , \nnr
S & = & \frac{\pi^2 [(r_+^2 + a^2)
(r_+^2 + b^2) + q r_+^2]}{2 \Xi_a \Xi_b r_+} .
\eea
In our asymptotically rotating coordinate frame, the angular velocities of
the horizon are
\ben
\hat{\Omega}_a = \frac{\Xi_a a (r_+^2 +
b^2)}{(r_+^2 + a^2) (r_+^2 + b^2) + q r_+^2} , \quad \hat{\Omega}_b =
\frac{\Xi_b b (r_+^2 + a^2)}{(r_+^2 + a^2) (r_+^2 + b^2) + q r_+^2} .
\een

For an extremal solution, with a horizon at $\hat{r} = r_0$,
we have $R'|_{\hat{r} = r_0} = 0$, and so
%%%%
\ben
r_0^4 - a^2 b^2 + g^2 r_0^4 (2 r_0^2 + a^2 + b^2 + 2 q) = 0 .
\een
%%%
Then we have the near-horizon expansion
%%%
\ben
R = V (\hat{r} - r_0)^2 + \textrm{O} (\hat{r} - r_0)^3 ,
\een
%%%
where
%%%
\ben
V = 1 - \frac{3 a^2 b^2}{r_0^4} + g^2 (6 r_0^2 + a^2 + b^2 + 2 q) .
\een
%%%
To obtain the near-horizon geometry, we make the coordinate changes
\ben
\hat{r} = r_0(1 + \lambda \rho)\,, \quad \hat{\phi}_1 = \phi_1 +
\hat{\Omega}_a^0 \hat{t} , \quad \hat{\phi}_2 = \phi_2 +
\hat{\Omega}_b^0 \hat{t} , \quad \hat{t} =
\frac{t}{2 \pi \dTH r_0 \lambda} ,
\een
%%%
and then take the limit $\lambda \rightarrow 0$.  The near-horizon
geometry is
%%%
\bea
\ud s^2 & = & H_0^{2/3} \bigg\{ \frac{r_0^2 + y^2}{V} \left(
- \rho^2 \ud t^2 + \frac{\ud \rho^2}{\rho^2} \right) +
\frac{r_0^2 + y^2}{Y} \ud y^2 \nnr
&& + \frac{Y}{r_0^2 + y^2} \left[ \frac{2 r_0}{H_0 V} \rho \,
\ud t + \frac{a (r_0^2 + a^2 + q)}{H_0 \Xi_a (a^2 - b^2)} \ud \phi_1 +
\frac{b (r_0^2 + b^2 + q)}{H_0 \Xi_b (b^2 - a^2)} \ud \phi_2 \right] ^2 \nnr
&& + \frac{a^2 b^2}{r_0^2 y^2} \bigg[ \frac{2}{H_0 r_0 V}
\left( r_0^2 + y^2 + \frac{q y^2}{r_0^2 + y^2} \right) \rho \,
\ud t + \frac{[(r_0^2 + a^2) (r_0^2 + y^2) + q r_0^2]
(a^2 - y^2)}{H_0 (r_0^2 + y^2) \Xi_a a (a^2 - b^2)} \ud \phi_1 \nnr
&& + \frac{[(r_0^2 + b^2) (r_0^2 + y^2) + q r_0^2]
(b^2 - y^2)}{H_0 (r_0^2 + y^2) \Xi_b b (b^2 - a^2)} \ud \phi_2
\bigg]^2 \bigg\} ,
\eea
%%%
where $H_0 = H | _{\hat{r} = r_0}$.  This can be cast in the form of
(\ref{dgenh}), so the Cardy formulae are satisfied.

For the extremal solution, the Frolov--Thorne temperatures are
%%%%
\ben
T_0 = 0 , \quad T_1 = \frac{V r_0 [(r_0^2 + a^2) (r_0^2 + b^2) +
q r_0^2]}{4 \pi \Xi_a a [(r_0^2 + b^2)^2 + q b^2]} ,
\quad T_2 = \frac{V r_0 [(r_0^2 + a^2) (r_0^2 + b^2) +
q r_0^2]}{4 \pi \Xi_b b [(r_0^2 + a^2)^2 + q a^2]} .
\een
The central charges are 
\ben
c_1 = \frac{6 \pi a [(r_0^2 + b^2)^2 + q b^2]}{V \Xi_b r_0^2} ,
\quad c_2 = \frac{6 \pi b [(r_0^2 + a^2)^2 + q a^2]}{V \Xi_a r_0^2} .
\een

\subsubsection{Equal angular momenta}

Charged rotating black holes with both angular momenta equal and
three arbitrary U(1) charges in $D=5$ gauged supergravity were
obtained in \cite{clpu1cube}.  Owing to the equality of the angular
momenta, the solution is of cohomogeneity-one.  The metric
has the form \cite{clpu1cube}
%%%
\be
ds_5^2 = -\fft{R\, Y}{f_1} d \hat t ^2 +
\fft{\hat r^2 R}{Y} d\hat r^2 +
\ft14 R (d\theta^2 + \sin^2\theta d\phi^2) +
\fft{f_1}{4R^2} \left( d\psi + \cos\theta\, d\phi -
2 \fft{f_2}{f_1} d\hat t\right) ^2
\,,
\ee
%%%
where $Y, R, f_1$ and $f_2$ are functions of the radial variable $\hat
r$ only, and were presented in detail in \cite{clpu1cube}.  The
angular coordinates $\phi$ and $\psi$ are related to the standard
$2\pi$-period azimuthal coordinates $\hat \phi_1$ and $\hat \phi_2$ as
follows:
%%%
\be
\phi =\hat \phi_1 - \hat \phi_2\,,\qquad
\psi =\hat \phi_1 + \hat \phi_2\,.
\ee
%%%
The extremal limit is achieved when $Y$ has a double root at
$\hat r=r_0$.  As in the previous cases, we make the
following coordinate transformation:
%%%
\be
\hat r = r_0(1 + \lambda \rho) \,,\qquad
\hat\phi_i = \phi_i + \fft{f_2(r_0)}{f_1(r_0)} \hat t\,,\qquad
\hat t= \fft{\sqrt{f_1(r_0)}}{V \lambda} t\,,
\ee
%%%%
where $V=\ft12 Y''(r_0)$.  Taking the $\lambda\rightarrow 0$ limit,
it is straightforward to obtain the near-horizon geometry, given by
%%%
\bea
ds_5^2&=&\fft{r_0^2 R(r_0)}{V} \left( -\rho^2 dt^2 + \fft{d\rho^2}{\rho^2}
\right) + \ft14 R(r_0) [ d\theta^2 + \sin^2 \theta \, 
(\tilde e_1-\tilde e_2)^2 ] \nn\\
&&+\fft{f_1(r_0)}{4R(r_0)^2} \Big( \tilde e_1 + \tilde e_2 +
\cos\theta \, (\tilde e_1 -\tilde e_2)\Big)^2\,,
\eea
where $\tilde e_i = d \phi_i + k_i \rho \, dt$.  This is exactly the
same form discussed in section 2, and hence the Cardy formula is
satisfied.

\section{Six and Seven Dimensions}

\subsection{Six-dimensional gauged supergravity}

We consider here the black hole solution of six-dimensional SU(2)
gauged supergravity \cite{chow6}, which has two independent angular
momenta and a single U(1) charge in the Cartan subgroup of the gauge
group.

The metric is
%%%
\bea
\ud s^2 & = & H^{1/2} \bigg[ - \frac{R}{H^2 U} \mathcal{A}^2 +
\frac{U}{R} \ud \hat{r}^2 +
\frac{(\hat{r}^2 + y^2) (y^2 - z^2)}{Y} \ud y^2 +
\frac{(\hat{r}^2 + z^2) (z^2 - y^2)}{Z} \ud z^2 \nnr
&& +
\frac{Y}{(\hat{r}^2 + y^2) (y^2 - z^2)} \bigg( \ud \hat{t} -
(\hat{r}^2 + a^2) (a^2 - z^2) \frac{\ud \hat{\phi}_1}{\epsilon_1} -
(\hat{r}^2 + b^2) (b^2 - z^2) \frac{\ud \hat{\phi}_2}{\epsilon_2} -
\frac{q \hat{r} \mathcal{A} }{H U} \bigg) ^2 \nnr
&& + \frac{Z}{(\hat{r}^2 + z^2) (z^2 - y^2)} \bigg( \! \ud \hat{t} -
(\hat{r}^2 \! + a^2) (a^2 -
y^2) \frac{\ud \hat{\phi}_1}{\epsilon_1} - (\hat{r}^2 \! +
b^2) (b^2 - y^2) \frac{\ud \hat{\phi}_2}{\epsilon_2} -
\frac{q \hat{r} \mathcal{A} }{H U} \! \bigg) ^2 \bigg] , \nnr
\eea
%%%
where
%%%
\bea
&& R = (\hat{r}^2 + a^2) (\hat{r}^2 + b^2) + g^2 [\hat{r}
(\hat{r}^2 + a^2) + q] [\hat{r} (\hat{r}^2 + b^2) + q]
- 2 m \hat{r} , \nnr
&& Y = - (1 - g^2 y^2) (a^2 - y^2) (b^2 - y^2) , \quad Z =
- (1 - g^2 z^2) (a^2 - z^2) (b^2 - z^2) , \nnr
&& U = (\hat{r}^2 + y^2) (\hat{r}^2 + z^2) , \quad \epsilon_1 =
\Xi_a a (a^2 - b^2) , \quad \epsilon_2 = \Xi_b b (b^2 - a^2) , \nnr
&& \Xi_a = 1 - a^2 g^2 , \quad \Xi_b = 1 - b^2 g^2 ,
\quad H = 1 + \frac{q \hat{r}}{U} ,
\quad q = 2 m s^2 , \quad s = \sinh \delta , \nnr
&& \mathcal{A} = \ud \hat{t} - (a^2 - y^2) (a^2 -
z^2) \frac{\ud \hat{\phi}_1}{\epsilon_1} - (b^2 - y^2)
(b^2 - z^2) \frac{\ud \hat{\phi}_2}{\epsilon_2} .
\eea
%%%
The coordinate changes $\hat{\phi}_1 \rightarrow \hat{\phi}_1 - a g^2
\hat{t}$ and $\hat{\phi}_2 \rightarrow \hat{\phi}_2 - b g^2 \hat{t}$
would give an asymptotically non-rotating coordinate frame.  The
Hawking temperature and entropy are \bea
%%%
T_{\textrm{H}} & = & \frac{R'|_{\hat{r} = r_+}}{4 \pi
[(r_+^2 + a^2) (r_+^2 + b^2) + q r_+]} \nnr
& = & \frac{2 (1 + g^2 r_+^2) r_+^2 (2 r_+^2 + a^2 + b^2)
- (1 - g^2 r_+^2) (r_+^2 + a^2) (r_+^2 + b^2) + 4 q g^2 r_+^3 -
q^2 g^2}{4 \pi r_+ [(r_+^2 + a^2) (r_+^2 + b^2) + q r_+]} , \nnr
S & = & \frac{2 \pi^2 [(r_+^2 + a^2) (r_+^2 + b^2) +
q r_+]}{3 \Xi_a \Xi_b} .
\eea
In our asymptotically rotating coordinate frame, the angular
velocities of the horizon are
\ben
\hat{\Omega}_a = \frac{\Xi_a a (r_+^2 + b^2)}{(r_+^2 + a^2)
(r_+^2 + b^2) + q r_+} , \quad \hat{\Omega}_b = \frac{\Xi_b b
(r_+^2 + a^2)}{(r_+^2 + a^2) (r_+^2 + b^2) + q r_+} .
\een

For an extremal solution, with a horizon at $\hat{r} = r_0$, we have
$R|_{\hat{r} = r_0} = 0$ and $R'|_{\hat{r} = r_0} = 0$, and so
%%%
\ben
3 r_0^4 + (a^2 + b^2) r_0^2 - a^2 b^2 + g^2 r_0^2 [5 r_0^4 +
3 (a^2 + b^2) r_0^2 + a^2 b^2] + 4 q g^2 r_0^3 - q^2 g^2 = 0 .
\een
%%%
Then we have the near-horizon expansion
%%%
\ben
R = V (\hat{r} - r_0)^2 + \textrm{O} (\hat{r} - r_0)^3 ,
\een
%%%
where
\ben
V = 6 r_0^2 + a^2 + b^2 + g^2 [15 r_0^4 + 6 (a^2 + b^2) r_0^2 +
6 q r_0 + a^2 b^2] .
\een
%%%
To obtain the near-horizon geometry, we make the coordinate changes
%%%
\ben
\hat{r} = r_0(1 + \lambda \rho)\,, \quad \hat{\phi}_1 = \phi_1 +
\hat{\Omega}_a^0 \hat{t} , \quad \hat{\phi}_2 = \phi_2 +
\hat{\Omega}_b^0 \hat{t} , \quad \hat{t} = \frac{t}{2
\pi \dTH r_0 \lambda} ,
\een
%%%%
and then take the limit $\lambda \rightarrow 0$.  The near-horizon
geometry is
\bea
\ud s^2 & = & H_0^{1/2}\bigg[ \frac{\widetilde{U}}{V}
\left( - \rho^2 \, \ud t^2 + \frac{\ud \rho^2}{\rho^2} \right) +
\frac{(r_0^2 + y^2) (y^2 - z^2)}{Y} \ud y^2 +
\frac{(r_0^2 + z^2) (z^2 - y^2)}{Z} \ud z^2 \nnr
&& + \frac{Y}{(r_0^2 + y^2) (y^2 - z^2)} \bigg(
\frac{2 r_0 (r_0^2 + z^2)}{V} \rho \, \ud t +
(r_0^2 + a^2) (a^2 - z^2) \frac{\ud \phi_1}{\epsilon_1} \nnr
&& + (r_0^2 + b^2) (b^2 - z^2) \frac{\ud \phi_2}{\epsilon_2} +
\frac{q r_0}{H_0 \widetilde{U}} \widetilde{\mathcal{A}} \bigg) ^2 +
\frac{Z}{(r_0^2 + z^2) (z^2 - y^2)} \bigg( \frac{2 r_0 (r_0^2 + y^2)}{V}
\rho \, \ud t \nnr
&& + (r_0^2 + a^2) (a^2 - y^2) \frac{\ud \phi_1}{\epsilon_1} +
(r_0^2 + b^2) (b^2 - y^2) \frac{\ud \phi_2}{\epsilon_2} +
\frac{q r_0}{H_0 \widetilde{U}} \widetilde{\mathcal{A}} \bigg) ^2 \bigg] ,
\eea
where $\widetilde{U} = U | _{\hat{r} = r_0}$, $H_0 = H | _{\hat{r} = r_0}$,
and
\ben
\widetilde{\mathcal{A}} = \frac{- 3 r_0^4 - r_0^2 (y^2 + z^2) +
y^2 z^2}{V r_0} \rho \, \ud t - (a^2 - y^2) (a^2 - z^2)
\frac{\ud \phi_1}{\epsilon_1} - (b^2 - y^2) (b^2 - z^2)
\frac{\ud \phi_2}{\epsilon_2} .
\een
%%%
This can be cast in the form of (\ref{dgenh}), so the Cardy formulae
are satisfied.

For an extremal solution, the Frolov--Thorne temperatures are
%%%
\ben
T_1 = \frac{V [(r_0^2 + a^2) (r_0^2 + b^2) +
q r_0]}{2 \pi \Xi_a a [2 r_0 (r_0^2 + b^2)^2 + q (b^2 - r_0^2)]} ,
\quad T_2 = \frac{V [(r_0^2 + a^2) (r_0^2 + b^2) + q r_0]}{2 \pi \Xi_b
b [2 r_0 (r_0^2 + a^2)^2 + q (a^2 - r_0^2)]} ,
\een
%%%
and $T_0 = 0$.  The central charges are 
%%%
\ben
c_1 = \frac{4 \pi a [2 r_0 (r_0^2 + b^2)^2 + q (b^2 - r_0^2)]}{V \Xi_b} ,
\quad c_2 = \frac{4 \pi b [2 r_0 (r_0^2 + a^2)^2 +
q (a^2 - r_0^2)]}{V \Xi_a} .
\een

\subsection{Seven-dimensional gauged supergravity}

We consider here the black hole solution of \cite{chow7} in
seven-dimensional SO(5) gauged supergravity.  It possesses three
independent angular momenta and a single charge parameter,
corresponding to two equal U(1) charges in the U(1)$^2$ Cartan
subgroup of the full gauge group.

The metric is
%%%%
\bea
\ud s^2 & = & H^{2/5} \bigg\{ - \frac{R}{H^2 U} \mathcal{A}^2 +
\frac{U}{R} \ud \hat{r}^2 + \frac{(\hat{r}^2 + y^2)(y^2 - z^2)}{Y}
\ud y^2 + \frac{(\hat{r}^2 + z^2) (z^2 - y^2)}{Z} \ud z^2 \nnr
&& + \frac{Y}{(\hat{r}^2 + y^2) (y^2 - z^2)} \bigg( \ud \hat{t} -
\sum_{i=1}^3 \frac{(\hat{r}^2 + a_i^2) \gamma_i}{a_i^2 - y^2}
\frac{\ud \hat{\phi}_i}{\epsilon_i} - \frac{q}{H U}
\mathcal{A} \bigg) ^2 \nnr
&& + \frac{Z}{(\hat{r}^2 + z^2) (z^2 - y^2)} \bigg( \ud \hat{t} -
\sum_{i=1}^3 \frac{(\hat{r}^2 + a_i^2) \gamma_i}{a_i^2 - z^2}
\frac{\ud \hat{\phi}_i}{\epsilon_i} - \frac{q}{H U} \mathcal{A}
\bigg) ^2 \nnr
&& + \frac{a_1^2 a_2^2 a_3^2}{\hat{r}^2 y^2 z^2} \bigg[ \ud \hat{t} -
\sum_{i=1}^3 \frac{(\hat{r}^2 + a_i^2) \gamma_i}{a_i^2}
\frac{\ud \hat{\phi}_i}{\epsilon_i} - \frac{q}{H U} \left( 1 +
\frac{g y^2 z^2}{a_1 a_2 a_3} \right) \mathcal{A} \bigg] ^2 \bigg\} ,
\eea
where
\bea
&& R = \frac{1 + g^2 \hat{r}^2}{\hat{r}^2} \prod_{i=1}^3 (\hat{r}^2 +
a_i^2) + q
g^2 ( 2 \hat{r}^2 + a_1^2 + a_2^2 + a_3^2 ) -
\frac{2 q g a_1 a_2 a_3}{\hat{r}^2} +
\frac{q^2 g^2}{\hat{r}^2} - 2m , \nnr
&& Y = \frac{1 - g^2 y^2}{y^2} \prod_{i=1}^3 (a_i^2 - y^2) , \quad Z =
\frac{1 - g^2 z^2}{z^2} \prod_{i=1}^3 (a_i^2 - z^2) ,
\quad U = (\hat{r}^2 + y^2) (\hat{r}^2 + z^2) , \nnr
&& \gamma_i = a_i^2 (a_i^2 - y^2) (a_i^2 - z^2) , \quad 
\epsilon_i = \Xi_i a_i \prod_{j \neq i} (a_i^2 - a_j^2) , \quad
\Xi_i = 1 - a_i^2 g^2 , \nnr
&& H = 1 + \frac{q}{(\hat{r}^2 + y^2) (\hat{r}^2 + z^2)} , \quad
q = 2 m s^2 , \quad s = \sinh \delta , \quad \mathcal{A} =
\ud \hat{t} - \sum_{i=1}^3 \gamma_i \frac{\ud \hat{\phi}_i}{\epsilon_i} .
\eea
%%%
The coordinate changes
$\hat{\phi}_i \rightarrow \hat{\phi}_i - a_i g^2 \hat{t}$ would give
an asymptotically non-rotating coordinate frame.  The Hawking
temperature and entropy are
%%%
\bea
T_{\textrm{H}} & = & \frac{(\hat{r}^2 R)' |_{\hat{r} = r_+}}{4 \pi
[(r_+^2 + a_1^2) (r_+^2 + a_2^2) (r_+^2 + a_3^2) + q (r_+^2 - a_1 a_2
a_3 g)]} \nnr
& = & \frac{(1 + g^2 r_+^2) r_+^2 \sum_i \prod_{j \neq i}
(r_+^2 + a_j^2) - \prod_i (r_+^2 + a_i^2)  +
2q ( g^2 r_+^4 + g a_1 a_2 a_3 ) - q^2 g^2}{2
  \pi r_+ [(r_+^2 + a_1^2) (r_+^2 + a_2^2) (r_+^2 +
  a_3^2) + q (r_+^2 - a_1 a_2 a_3 g)]}, \nnr
S & = & \frac{\pi^3 [(r_+^2 + a_1^2) (r_+^2 + a_2^2)
(r_+^2 + a_3^2) + q (r_+^2 - a_1 a_2 a_3 g)]}{4 \Xi_1 \Xi_2 \Xi_3 r_+} .
\eea
In our asymptotically rotating coordinate frame, the angular
velocities of the horizon are
\ben
\hat{\Omega}_i = \frac{\Xi_i [a_i \prod_{j \neq i}
(r_+^2 + a_j^2) - q \prod_{j \neq i} a_j g]}{(r_+^2
  + a_1^2) (r_+^2 + a_2^2) (r_+^2 + a_3^2) + q (r_+^2 - a_1 a_2
  a_3 g)} .
\een

For an extremal solution, with a horizon at $\hat{r} = r_0$, we have
$R'|_{\hat{r} = r_0} = 0$, and so
%%%
\bea
&& 2 r_0^6 + (a_1^2 + a_2^2 + a_3^2) r_0^4 - a_1^2 a_2^2 a_3^2 +
g^2 [3 r_0^8 + 2 (a_1^2 + a_2^2 + a_3^2) r_0^6 \nnr
&& + (a_1^2 a_2^2 + a_2^2 a_3^2 + a_3^2 a_1^2 + 2 q) r_0^4 - q^2]
+ 2 q g a_1 a_2 a_3 = 0 . 
\eea
%%%
Then we have the near-horizon expansion
%%%
\ben
R = V (\hat{r} - r_0)^2 + \textrm{O} (\hat{r} - r_0)^3 ,
\een
%%%
where
%%%
\bea
V & = & 6 r_0^2 + a_1^2 + a_2^2 + a_3^2 +
\frac{3 (a_1 a_2 a_3 - q g)^2}{r_0^4} \nnr
&& + g^2 [15 r_0^4 + 6 (a_1^2 + a_2^2 + a_3^2) r_0^2 +
a_1^2 a_2^2 + a_2^2 a_3^2 + a_3^2 a_1^2 + 2 q] .
\eea
%%%%
To obtain the near-horizon geometry, we make the coordinate changes
%%%
\ben
\hat{r} = r_0 (1 + \lambda \rho) , \quad \hat{\phi}_i =
\phi_i + \Omega_i^0 \hat{t} , \quad \hat{t} =
\frac{t}{2 \pi \dTH r_0 \lambda} ,
\een
%%%
and then take the limit $\lambda \rightarrow 0$.  The near-horizon
geometry is
%%%%
{\crampest
\bea
\ud s^2 & = & H_0^{2/5} \bigg\{ \frac{\widetilde{U}}{V}
\left( - \rho^2 \ud t^2 + \frac{\ud \rho^2}{\rho^2} \right) +
\frac{(r_0^2 + y^2) (y^2 - z^2)}{Y} \ud y^2 + \frac{(r_0^2 + z^2)
(z^2 - y^2)}{Z} \ud z^2 \nnr
%%%%
&& + \frac{Y}{(r_0^2 + y^2) (y^2 - z^2)} \bigg( \frac{2 r_0
(r_0^2 + z^2)}{V} \rho \, \ud t + \sum_{i=1}^3
\frac{(r_0^2 + a_i^2) \gamma_i}{a_i^2 - y^2}
\frac{\ud \phi_i}{\epsilon_i} +
\frac{q}{H_0 \widetilde{U}} \widetilde{\mathcal{A}} \bigg) ^2 \nnr
&& + \frac{Z}{(r_0^2 + z^2) (z^2 - y^2)}
\bigg( \frac{2 r_0 (r_0^2 + y^2)}{V} \rho \,
\ud t + \sum_{i=1}^3 \frac{(r_0^2 + a_i^2) \gamma_i}{a_i^2 - z^2}
\frac{\ud \phi_i}{\epsilon_i} + \frac{q}{H_0 \widetilde{U}}
\widetilde{\mathcal{A}} \bigg) ^2\\
&& + \frac{a_1^2 a_2^2 a_3^2}{r_0^2 y^2 z^2}
\bigg[ \frac{2}{V r_0} \left( \widetilde{U} - \frac{q g y^2 z^2}{a_1
      a_2 a_3} \right) \rho \, \ud t +
\sum_{i=1}^3 \frac{(r_0^2 + a_i^2) \gamma_i}{a_i^2}
\frac{\ud \phi_i}{\epsilon_i} + \frac{q}{H_0 \widetilde{U}}
\left( 1 + \frac{g y^2 z^2}{a_1 a_2 a_3} \right)
\widetilde{\mathcal{A}} \bigg ] ^2 \bigg\} ,\nn
\eea}
%%%
where $\widetilde{U} = U | _{\hat{r} = r_0}$, $H_0 = H| _{\hat{r} =
r_0}$, and
%%%
\ben
\widetilde{\mathcal{A}} = - \frac{2 r_0 (2 r_0^2 + y^2 + z^2)}{V}
\rho \, \ud t - \sum_{i=1}^3 \gamma_i \frac{\ud \phi_i}{\epsilon_i} .
\een
%%%
This can be cast in the form of (\ref{dgenh}), and so the Cardy
formulae are satisfied.

For an extremal solution, the Frolov--Thorne temperatures are
\bea
T_1 & = & \frac{V r_0 [(r_0^2 + a_1^2) (r_0^2 + a_2^2) (r_0^2 + a_3^2)
+ q  (r_0^2 - a_1 a_2 a_3 g)]}{4 \pi \Xi_1} [ a_1 (r_0^2 + a_2^2)^2
  (r_0^2 + a_3^2)^2 \nnr
&& + q a_1 (a_2^2 a_3^2 - r_0^4) - q g a_2 a_3 (3
  r_0^4 + 2 a_2^2 r_0^2 + 2 a_3^2 r_0^2 + a_2^2 a_3^2 ) - q^2 g a_2
  a_3 ] ^{-1} , 
\eea
temperatures $T_2$ and $T_3$, obtained by cyclic permutation of $a_i$,
$i = 1, 2, 3$, and also $T_0 = 0$.  The central charges are 
\bea
c_1 & = & \frac{3 \pi^2}{\Xi_2 \Xi_3 V r_0^2} [ a_1 (r_0^2 + a_2^2)^2
  (r_0^2 + a_3^2)^2 + q a_1 (a_2^2 a_3^2 - r_0^4) \nnr
&& - q g a_2 a_3 (3
  r_0^4 + 2 a_2^2 r_0^2 + 2 a_3^2 r_0^2 + a_2^2 a_3^2 ) - q^2 g a_2
  a_3 ] ,
\eea
and also $c_2$ and $c_3$, obtained by cyclic permutation of $a_i$, $i
= 1, 2, 3$.

\section{Arbitrary Dimensions}

\subsection{Higher-dimensional Kerr--AdS}

The extremal black hole/CFT correspondence for the higher-dimensional
Kerr--AdS solution \cite{gilupapo1, gilupapo2} was previously
considered in \cite{lumeipope}, where it was shown that the Cardy
formulae are satisfied.  We return to this example, showing directly
that the near-horizon geometry of its extremal limit can be cast in
the form of (\ref{dgenh}).  (Note that the near-horizon geometry of
the extremal Kerr--AdS black hole in $D=5$ was obtained in
\cite{kun2}, and that of the Myers--Perry solution in
\cite{kun3}.)

\subsubsection{Even dimensions $D = 2n$}

The Kerr--AdS metric in even dimensions $D = 2n$ is
\ben
\ud s ^2 = - \frac{R}{U} \mathcal{A}^2 +
\frac{U}{R} \ud \hat{r}^2 + \sum_{\alpha = 1}^{n-1}
\frac{U_\alpha}{X_\alpha} \ud y_\alpha^2 + \sum_{\alpha = 1}^{n-1}
\frac{X_\alpha}{U_\alpha}
\bigg( \ud \hat{t} - \sum_{i=1}^{n-1} \frac{(\hat{r}^2 + a_i^2)
\gamma_i}{a_i^2 - y_\alpha^2} \frac{\ud \hat{\phi}_i}{\epsilon_i}
\bigg) ^2 ,
\een
where
\bea
&& R = \prod_{k=1}^{n-1} (\hat{r}^2 + a_k^2) - 2 m \hat{r} ,
\quad X_\alpha = - \prod_{k=1}^{n-1} (a_k^2 - y_\alpha^2) ,
\quad U = \prod_{\alpha = 1}^{n-1} (\hat{r}^2 + y_\alpha^2) ,
\nnr
&& U_\alpha = - (\hat{r}^2 + y_\alpha^2) \sideset{}{'}
\prod_{\beta = 1}^{n-1} (y_\beta^2 - y_\alpha^2) ,
\quad \gamma_i = \prod_{\alpha = 1}^{n-1} (a_i^2 - y_\alpha^2) ,
\quad \epsilon_i = \Xi_i a_i \sideset{}{'} \prod_{k=1}^{n-1}
(a_i^2 - a_k^2) , \nnr
&& \Xi_i = 1 - a_i^2 g^2 , \quad \mathcal{A} = \ud \hat{t} -
\sum_{i=1}^{n-1} \gamma_i \frac{\ud \hat{\phi}_i}{\epsilon_i} .
\eea
The notation $\prod '$ means that we omit the factor that vanishes
from a product.

For the extremal solution, the Frolov--Thorne temperatures are
\ben
T_i = \frac{V (r_0^2 + a_i^2)}{4 \pi \Xi_i a_i r_0
\prod_{j \neq i} (r_0^2 + a_j^2)} ,
\een
where $V = \frac{1}{2} R'' | _{\hat{r} =
r_0}$, and also $T_0 = 0$.  The near-horizon geometry is \cite{lumeipope}
\bea
\ud s^2 & = & \frac{\widetilde{U}}{V}
\left( - \rho^2 \ud t^2 + \frac{\ud \rho^2}{\rho^2} \right) +
\sum_{\alpha = 1}^{n-1} \frac{\widetilde{U}_\alpha}{X_\alpha}
\ud y_\alpha^2 \nnr
&& + \sum_{\alpha = 1}^{n-1} \frac{X_\alpha}{U_\alpha}
\bigg( \frac{2 r_0 \widetilde{U}}{V (r_0^2 + y_\alpha^2)}
\rho \, \ud t + \sum_{i=1}^{n-1} \frac{(r_0^2 + a_i^2)
\gamma_i}{a_i^2 - y_\alpha^2} \frac{\ud \phi_i}{\epsilon_i} \bigg) ^2 ,
\eea
%%%
where $\widetilde{U} = U |_{\hat{r} = r_0}$ and $\widetilde{U}_\alpha =
U_\alpha |_{\hat{r} = r_0}$.

We can cast this near-horizon geometry in the form of (\ref{dgenh}),
reading off $\tilde{g}_{ij}$ from $g_{\phi_i \phi_j}$.  To explicitly
see how for this example, we need to account for the correct
coefficients of $\ud t$ within the vielbeins.  From the partial
fraction decomposition
%%%
\ben
\frac{\widetilde{U}}{(r_0^2 + y_\alpha^2) \prod_{k=1}^{n-1}
(r_0^2 + a_k^2)} = \sum_{i=1}^{n-1} \frac{\Xi_i a_i \gamma_i}{
\epsilon_i (a_i^2 - y_\alpha^2) (r_0^2 + a_i^2)} ,
\een
%%%%
we see that this is indeed the case.  It follows that the Cardy
formulae are satisfied.

\subsubsection{Odd dimensions $D = 2n + 1$}

The Kerr--AdS metric in odd dimensions $D = 2n + 1$ is
\bea
\ud s ^2 & = & - \frac{R}{U} \mathcal{A}^2 +
\frac{U}{R} \ud \hat{r}^2 + \sum_{\alpha = 1}^{n-1}
\frac{U_\alpha}{X_\alpha} \ud y_\alpha^2 + \sum_{\alpha = 1}^{n-1}
\frac{X_\alpha}{U_\alpha}
\bigg( \ud \hat{t} - \sum_{i=1}^n \frac{(\hat{r}^2 + a_i^2)
\gamma_i}{a_i^2 - y_\alpha^2} \frac{\ud \hat{\phi}_i}{\epsilon_i}
\bigg) ^2 \nnr
&& + \frac{\prod_{k=1}^n a_k^2}{r^2 \prod_{\alpha = 1}^{n-1}
y_\alpha^2} \bigg( \ud \hat{t} - \sum_{i=1}^n \frac{(\hat{r}^2 + a_i^2)
\gamma_i}{a_i^2} \frac{\ud \hat{\phi}_i}{\epsilon_i} \bigg) ^2 ,
\eea
where
\bea
&& R = \frac{1}{\hat{r}^2}
\prod_{k=1}^n (\hat{r}^2 + a_k^2) - 2 m , \quad X_\alpha =
\frac{1}{y_\alpha^2} \prod_{k=1}^n (a_k^2 - y_\alpha^2) ,
\quad U = \prod_{\alpha = 1}^{n-1} (\hat{r}^2 + y_\alpha^2) ,
\nnr
&& U_\alpha = - (\hat{r}^2 + y_\alpha^2) \sideset{}{'}
\prod_{\beta = 1}^{n-1} (y_\beta^2 - y_\alpha^2) ,
\quad \gamma_i = a_i^2 \prod_{\alpha = 1}^{n-1}
(a_i^2 - y_\alpha^2) , \quad \epsilon_i = \Xi_i a_i \sideset{}{'}
\prod_{k=1}^n (a_i^2 - a_k^2) ,
\nnr
&& \Xi_i = 1 - a_i^2 g^2 , \quad \mathcal{A} = \ud \hat{t} -
\sum_{i=1}^n \gamma_i \frac{\ud \hat{\phi}_i}{\epsilon_i} .
\eea

For the extremal solution, the Frolov--Thorne temperatures are
\ben
T_i = \frac{V r_0 (r_0^2 + a_i^2)}{4 \pi
\Xi_i a_i \prod_{j \neq i} (r_0^2 + a_j^2)} .
\een
where $V = \frac{1}{2} R'' | _{\hat{r} = r_0}$, and also $T_0 = 0$.  The near-horizon geometry is \cite{lumeipope}
\bea
\ud s^2 & = & \frac{\widetilde{U}}{V}
\left( - \rho^2 \ud t^2 + \frac{\ud \rho^2}{\rho^2} \right) +
\sum_{\alpha = 1}^{n-1} \frac{\widetilde{U}_\alpha}{X_\alpha}
\ud y_\alpha^2 \nnr
&& + \sum_{\alpha = 1}^{n-1} \frac{X_\alpha}{U_\alpha} \bigg(
\frac{2 r_0 \widetilde{U}}{V (r_0^2 + y_\alpha^2)} \rho \, \ud t +
\sum_{i=1}^n \frac{(r_0^2 + a_i^2) \gamma_i}{a_i^2 - y_\alpha^2}
\frac{\ud \phi_i}{\epsilon_i} \bigg) ^2 \nnr
&& + \frac{\prod_{k=1}^n a_k^2}{r^2 \prod_{\alpha = 1}^{n-1}
y_\alpha^2} \bigg( \frac{2 \widetilde{U}}{V r_0} \rho \, \ud t
+ \sum_{i=1}^n \frac{(r_0^2 + a_i^2) \gamma_i}{a_i^2}
\frac{\ud \phi_i}{\epsilon_i} \bigg) ^2 ,
\eea
where $\widetilde{U} = U |_{\hat{r} = r_0}$ and $\widetilde{U}_\alpha
= U_\alpha |_{\hat{r} = r_0}$.

Analogously to the even-dimensional case, the near-horizon geometry
can be cast in the form of (\ref{dgenh}).  The analogous partial
fraction decomposition used is
%%%
\ben
\frac{r_0^2 \widetilde{U}}{(r_0^2 + y_\alpha^2)
\prod_{k=1}^n (r_0^2 + a_k^2)} = \sum_{i=1}^n
\frac{\Xi_i a_i \gamma_i}{\epsilon_i (a_i^2 - y_\alpha^2) (r_0^2 + a_i^2)} .
\een
It again follows that the Cardy formulae are satisfied.

\subsection{Charged rotating black holes in ungauged supergravity}

The solution considered here is the two-charge Cveti\v{c}--Youm
solution \cite{cvetyoum2}, with the simplification of \cite{chow7}
that both charges are equal.  It can be regarded as a solution of
toroidally compactified heterotic supergravity in dimension $4 \leq D
\leq 9$, although the construction generalizes to arbitrary dimension
as a solution of a low-energy effective action of bosonic strings.
This solution underlies the ungauged limit of some of the gauged black
hole solutions that we have considered above.  We use the form of the
metric in \cite{chow7}.

\subsubsection{Even dimensions $D = 2n$}

In even dimensions $D = 2n$, the metric is
\bea
\ud s ^2 & = & H^{2/(D-2)} \bigg[ - \frac{R}{H^2 U} \mathcal{A}^2 +
\frac{U}{R} \ud \hat{r}^2 + \sum_{\alpha = 1}^{n-1}
\frac{U_\alpha}{X_\alpha} \ud y_\alpha^2 \nnr
&& + \sum_{\alpha = 1}^{n-1} \frac{X_\alpha}{U_\alpha}
\bigg( \ud \hat{t} - \sum_{i=1}^{n-1} \frac{(\hat{r}^2 + a_i^2)
\gamma_i}{a_i^2 - y_\alpha^2} \frac{\ud \hat{\phi}_i}{\epsilon_i}
- \frac{q \hat{r}}{H U} \mathcal{A} \bigg) ^2 \bigg] ,
\eea
%%%
where
%%%
\bea
&& R = \prod_{k=1}^{n-1} (\hat{r}^2 + a_k^2) - 2 m \hat{r} ,
\quad X_\alpha = - \prod_{k=1}^{n-1} (a_k^2 - y_\alpha^2) ,
\quad U = \prod_{\alpha = 1}^{n-1} (\hat{r}^2 + y_\alpha^2) ,
\nnr
&& U_\alpha = - (\hat{r}^2 + y_\alpha^2) \sideset{}{'}
\prod_{\beta = 1}^{n-1} (y_\beta^2 - y_\alpha^2) ,
\quad \gamma_i = \prod_{\alpha = 1}^{n-1} (a_i^2 - y_\alpha^2) , \quad
\epsilon_i = a_i \sideset{}{'} \prod_{k=1}^{n-1} (a_i^2 - a_k^2) ,\nnr
&& H = 1 + \frac{q \hat{r}}{U} , \quad q = 2 m s^2 ,
\quad s = \sinh \delta , \quad \mathcal{A} = \ud \hat{t} -
\sum_{i=1}^{n-1} \gamma_i \frac{\ud \hat{\phi}_i}{\epsilon_i} .
\eea
%%%%
The Hawking temperature and entropy are
%%%%
\ben
T_{\textrm{H}} =
\frac{R' | _{\hat{r} = r_+}}{4 \pi [\prod_{k=1}^{n-1} (r_+^2 + a_k^2)
+ q r_+]} , \quad S = \frac{\mathcal{A}_{D-2} [\prod_{k=1}^{n-1}
(r_+^2 + a_k^2) + q r_+]}{4 r_+} ,
\een
%%%%
where $\mathcal{A}_{D-2} = 2 \pi^{(D-1)/2} / \Gamma[(D-1)/2]$ is the
volume of a unit $(D-2)$-sphere, so for example $\mathcal{A}_2 = 4 \pi$ and
$\mathcal{A}_4 = \frac{8}{3} \pi^2$.  The angular velocities of the
horizon are
%%%%
\ben
\Omega_i = \frac{a_i \prod_{j \neq i} (r_+^2 + a_j^2)}{
\prod_{k=1}^{n-1} (r_+^2 + a_k^2) + q r_+} .
\een

For an extremal solution, with a horizon at $\hat{r} = r_0$, we have
$R|_{\hat{r} = r_0} = 0$ and $R'|_{\hat{r} = r_0} = 0$, and so
%%%
\ben
\sum_{i=1}^{n-1} \frac{1}{r_0^2 + a_i^2} = \frac{1}{2 r_0^2} .
\label{extremeeven}
\een
Then we have the near-horizon expansion
\ben
%%%
R = V (\hat{r} - r_0)^2 + \textrm{O} (\hat{r} - r_0)^3 ,
\quad V = \tfrac{1}{2} R'' |_{\hat{r} = r_0} .
\een
Since $R'|_{\hat{r} = r_0} = 0$ for an extremal solution, we have
\ben
\dTH = \frac{V}{2 \pi [\prod_{k=1}^{n-1} (r_0^2 + a_k^2) + q r_0]} ,
\een
and, using (\ref{extremeeven}), we obtain
\ben
\dOi = - \frac{2 a_i r_0 \prod_{j \neq i} (r_0^2 + a_j^2)}{
(r_0^2 + a_i^2) [\prod_{k=1}^{n-1} (r_0^2 + a_k^2) + q r_0]} .
\een
Therefore the Frolov--Thorne temperatures are
\ben
T_0 = 0 , \quad T_i = \frac{V (r_0^2 + a_i^2)}{4 \pi a_i r_0
\prod_{j \neq i} (r_0^2 + a_j^2)} .
\een

To obtain the near-horizon geometry, we make the coordinate changes
\ben
\hat{r} = r_0(1 + \lambda \rho) , \quad \hat{\phi}_i = \phi_i +
\Omega_i^0 \hat{t} , \quad \hat{t} = \frac{t}{2 \pi \dTH r_0 \lambda} ,
\een
and then take the limit $\lambda \rightarrow 0$.  The near-horizon
geometry is
\bea
\ud s^2 & = & H_0^{2/(D-2)} \bigg[ \frac{\widetilde{U}}{V}
\left( - \rho^2 \ud t^2 + \frac{\ud \rho^2}{\rho^2} \right) +
\sum_{\alpha = 1}^{n-1} \frac{\widetilde{U}_\alpha}{X_\alpha}
\ud y_\alpha^2 \nnr
&& + \sum_{\alpha = 1}^{n-1} \frac{X_\alpha}{U_\alpha}
\bigg( \frac{2 r_0 \widetilde{U}}{V (r_0^2 + y_\alpha^2)}
\rho \, \ud t + \sum_{i=1}^{n-1} \frac{(r_0^2 + a_i^2)
\gamma_i}{a_i^2 - y_\alpha^2} \frac{\ud \phi_i}{\epsilon_i} +
\frac{q r_0}{H_0 \widetilde{U}} \widetilde{\mathcal{A}}
\bigg) ^2 \bigg] ,
\eea
where $\widetilde{U} = U |_{\hat{r} = r_0}$, $\widetilde{U}_\alpha
= U_\alpha |_{\hat{r} = r_0}$, $H_0 = H | _{\hat{r} = r_0}$, and 
\ben
H'_0 = \frac{\pd H_0}{\pd r_0} = \frac{q r_0}{\widetilde{U}}
\bigg( \frac{1}{r_0} - \sum_{\alpha = 1}^{n-1}
\frac{2 r_0}{r_0^2 + y_\alpha^2} \bigg) , \quad
\widetilde{\mathcal{A}} = \frac{H'_0 \widetilde{U}^2}{V q r_0}
\rho \, \ud t - \sum_{i=1}^{n-1} \gamma_i \frac{\ud \phi_i}{\epsilon_i} .
\een

By checking $\ud t$ coefficients within the vielbeins, we can directly
see that this near-horizon geometry may be cast in the form of
(\ref{dgenh}).  Some terms follow in the same way as for the
higher-dimensional Kerr--AdS solution.  There are also extra terms
when charge is included; these extra terms are within
$\widetilde{\mathcal{A}}$.  To check these extra terms, we use the
identity
%%%%
\ben
\frac{1}{\widetilde{U}} \sum_{i=1}^{n-1} \frac{a_i \gamma_i
\prod_{j \neq i} (r_0^2 + a_j^2)}{\epsilon_i (r_0^2 + a_i^2)} =
\sum_{\alpha = 1}^{n-1} \frac{1}{r_0^2 + y_\alpha^2} -
\sum_{i=1}^{n-1} \frac{1}{r_0^2 + a_i^2} .
\een
%%%%
which is seen to hold by a partial fraction decomposition of the
entire left hand side.  On the right hand side, the coefficients of
$1/(r_0^2 + a_i^2)$ are trivial, and the coefficients of $1/(r_0^2 +
y_\alpha^2)$ in turn follow from the partial fraction decomposition
%%%%
\ben
\frac{U_\alpha}{(r_0^2 + y_\alpha^2) X_\alpha} = \sum_{i=1}^{n-1}
\frac{a_i \gamma_i}{\epsilon_i (a_i^2 - y_\alpha^2)} .
\een
%%%%
Now using the extremality condition (\ref{extremeeven}), we see that
%%%%
\ben
\frac{2 r_0}{\widetilde{U}} \sum_{i=1}^{n-1} \frac{a_i \gamma_i
\prod_{j \neq i} (r_0^2 + a_j^2)}{\epsilon_i (r_0^2 + a_i^2)} =
- \bigg( \frac{1}{r_0} - \sum_{\alpha = 1}^{n-1}
\frac{2 r_0}{r_0^2 + y_\alpha^2} \bigg) = -
\frac{H'_0 \widetilde{U}}{q r_0} ,
\een
%%%%
completing the verification.  It follows that the Cardy formulae are
satisfied.  The central charges are
%%%
\ben
c_i = \frac{3 \mathcal{A}_{D-2} a_i r_0 \prod_{j \neq i} (r_0^2 + a_j^2)
[\prod_{k=1}^{n-1} (r_0^2 + a_k^2) + q r_0]}{\pi V (r_0^2 + a_i^2)} .
\een

\subsubsection{Odd dimensions $D = 2n + 1$}

In odd dimensions $D = 2n + 1$, the metric is
\bea
\ud s ^2 & = & H^{2/(D-2)} \bigg[ - \frac{R}{H^2 U} \mathcal{A}^2 +
\frac{U}{R} \ud \hat{r}^2 + \sum_{\alpha = 1}^{n-1}
\frac{U_\alpha}{X_\alpha} \ud y_\alpha^2 \nnr
&& + \sum_{\alpha = 1}^{n-1} \frac{X_\alpha}{U_\alpha}
\bigg( \ud \hat{t} - \sum_{i=1}^n \frac{(\hat{r}^2 + a_i^2)
\gamma_i}{a_i^2 - y_\alpha^2} \frac{\ud \hat{\phi}_i}{\epsilon_i}
- \frac{q}{H U} \mathcal{A} \bigg) ^2 \nnr
&& + \frac{\prod_{k=1}^n a_k^2}{r^2 \prod_{\alpha = 1}^{n-1}
y_\alpha^2} \bigg( \ud \hat{t} - \sum_{i=1}^n \frac{(\hat{r}^2 + a_i^2)
\gamma_i}{a_i^2} \frac{\ud \hat{\phi}_i}{\epsilon_i} -
\frac{q}{H U} \mathcal{A} \bigg) ^2 \bigg] ,
\eea
where
\bea
&& R = \frac{1}{\hat{r}^2}
\prod_{k=1}^n (\hat{r}^2 + a_k^2) - 2 m , \quad X_\alpha =
\frac{1}{y_\alpha^2} \prod_{k=1}^n (a_k^2 - y_\alpha^2) ,
\quad U = \prod_{\alpha = 1}^{n-1} (\hat{r}^2 + y_\alpha^2) ,
\nnr
&& U_\alpha = - (\hat{r}^2 + y_\alpha^2) \sideset{}{'}
\prod_{\beta = 1}^{n-1} (y_\beta^2 - y_\alpha^2) ,
\quad \gamma_i = a_i^2 \prod_{\alpha = 1}^{n-1}
(a_i^2 - y_\alpha^2) , \quad \epsilon_i = a_i \sideset{}{'}
\prod_{k=1}^n (a_i^2 - a_k^2) ,
\nnr
&& H = 1 + \frac{q}{U} , \quad q = 2 m s^2 ,
\quad s = \sinh \delta , \quad \mathcal{A} = \ud \hat{t} -
\sum_{i=1}^n \gamma_i \frac{\ud \hat{\phi}_i}{\epsilon_i} .
\eea
%%%
The Hawking temperature and entropy are
%%%
\ben
T_{\textrm{H}} = \frac{(\hat{r}^2 R)' | _{\hat{r} = r_+}}{
4 \pi [\prod_{k=1}^n (r_+^2 + a_k^2) + q r_+^2]} ,
\quad S = \frac{\mathcal{A}_{D-2} [\prod_{k=1}^n (r_+^2 + a_k^2)
+ q r_+^2]}{4 r_+} ,
\een
%%%
where $\mathcal{A}_{D-2} = 2 \pi^{(D-1)/2} / \Gamma[(D-1)/2]$ is the
volume of a unit $(D-2)$-sphere, so for example $\mathcal{A}_3 = 2 \pi^2$ and
$\mathcal{A}_5 = \pi^3$.  The angular velocities of the horizon are
%%%
\ben
\Omega_i = \frac{a_i \prod_{j \neq i} (r_+^2 + a_j^2)}{\prod_{k=1}^n
(r_+^2 + a_k^2) + q r_+^2} .
\een

For an extremal solution, with a horizon at $\hat{r} = r_0$, we have
$R'|_{\hat{r} = r_0} = 0$, and so
%%%%
\ben
\sum_{i=1}^n \frac{1}{r_0^2 + a_i^2} = \frac{1}{r_0^2} .
\label{extremeodd}
\een
Then we have the near-horizon expansion
\ben
R = V (\hat{r} - r_0)^2 + \textrm{O} (\hat{r} - r_0)^3 ,
\quad V = \tfrac{1}{2} R'' |_{\hat{r} = r_0} .
\een
Since $R'|_{\hat{r} = r_0} = 0$ for an extremal solution, we have
\ben
\dTH = \frac{r_0^2 V}{2 \pi [\prod_{k=1}^n (r_0^2 + a_k^2) + q r_0^2]} ,
\een
and, using (\ref{extremeodd}), we obtain
\ben
\dOi = - \frac{2 a_i r_0 \prod_{j \neq i} (r_0^2 + a_j^2)}{
(r_0^2 + a_i^2) [\prod_{k=1}^n (r_0^2 + a_k^2) + q r_0^2]} , 
\een
Therefore the Frolov--Thorne temperatures are
\ben
T_0 = 0 , \quad T_i = \frac{V r_0 (r_0^2 + a_i^2)}{4 \pi
a_i \prod_{j \neq i} (r_0^2 + a_j^2)} .
\een

To obtain the near-horizon geometry, we make the coordinate changes
\ben
\hat{r} = r_0(1 + \lambda \rho) , \quad \hat{\phi}_i = \phi_i +
\Omega_i^0 \hat{t} , \quad \hat{t} = \frac{t}{2 \pi \dTH r_0 \lambda} ,
\een
%%%
and then take the limit $\lambda \rightarrow 0$.  The near-horizon
geometry is
%%%
\bea
\ud s^2 & = & H_0^{2/(D-2)} \bigg[ \frac{\widetilde{U}}{V}
\left( - \rho^2 \ud t^2 + \frac{\ud \rho^2}{\rho^2} \right) +
\sum_{\alpha = 1}^{n-1} \frac{\widetilde{U}_\alpha}{X_\alpha}
\ud y_\alpha^2 \nnr
&& + \sum_{\alpha = 1}^{n-1} \frac{X_\alpha}{U_\alpha} \bigg(
\frac{2 r_0 \widetilde{U}}{V (r_0^2 + y_\alpha^2)} \rho \, \ud t +
\sum_{i=1}^n \frac{(r_0^2 + a_i^2) \gamma_i}{a_i^2 - y_\alpha^2}
\frac{\ud \phi_i}{\epsilon_i} + \frac{q}{H_0 \widetilde{U}}
\widetilde{\mathcal{A}} \bigg) ^2 \nnr
&& + \frac{\prod_{k=1}^n a_k^2}{r^2 \prod_{\alpha = 1}^{n-1}
y_\alpha^2} \bigg( \frac{2 \widetilde{U}}{V r_0} \rho \, \ud t
+ \sum_{i=1}^n \frac{(r_0^2 + a_i^2) \gamma_i}{a_i^2}
\frac{\ud \phi_i}{\epsilon_i} + \frac{q}{H_0 \widetilde{U}}
\widetilde{\mathcal{A}} \bigg) ^2 \bigg] ,
\eea
where $\widetilde{U} = U |_{\hat{r} = r_0}$, $\widetilde{U}_\alpha
= U_\alpha |_{\hat{r} = r_0}$, $H_0 = H | _{\hat{r} = r_0}$, and
\ben
H'_0 = \frac{\pd H_0}{\pd r_0} = - \frac{q}{\widetilde{U}}
\sum_{\alpha = 1}^{n-1} \frac{2 r_0}{r_0^2 + y_\alpha^2} ,
\quad \widetilde{\mathcal{A}} = \frac{H'_0 \widetilde{U}^2}{V q}
\rho \, \ud t - \sum_{i=1}^n \gamma_i \frac{\ud \phi_i}{\epsilon_i} .
\een
As in the even-dimensional case, we can directly see that this
near-horizon geometry can be cast in the form of (\ref{dgenh}) by
checking $\ud t$ coefficients.  The analogous identities needed are
\ben
\frac{1}{r_0^2 \widetilde{U}} \sum_{i=1}^n \frac{a_i \gamma_i
\prod_{j \neq i} (r_0^2 + a_j^2)}{\epsilon_i (r_0^2 + a_i^2)} =
\sum_{\alpha = 1}^{n-1} \frac{1}{r_0^2 + y_\alpha^2} -
\sum_{i=1}^n \frac{1}{r_0^2 + a_i^2} + \frac{1}{r_0^2} ,
\label{oddsum}
\een
\ben
\frac{U_\alpha}{(r_0^2 + y_\alpha^2) X_\alpha} = \sum_{i=1}^n
\frac{\gamma_i}{a_i \epsilon_i (a_i^2 - y_\alpha^2)} .
\een
%%%
The $1/r_0^2$ coefficient on the right hand side of (\ref{oddsum})
follows from the identity
%%%
\ben
\sum_{i=1}^n \frac{\gamma_i}{a_i^3 \epsilon_i} =
\frac{\prod_{\alpha = 1}^{n-1} y_\alpha^2}{\prod_{k=1}^n a_k^2} ,
\een
%%%
as seen by a partial fraction decomposition of one of the terms on the
left hand side.  Now using the extremality condition
(\ref{extremeodd}), we see that
%%%%
\ben
\frac{2}{r_0 \widetilde{U}} \sum_{i=1}^n \frac{a_i
\gamma_i \prod_{j \neq i} (r_0^2 + a_j^2)}{\epsilon_i (r_0^2 + a_i^2)}
= \sum_{\alpha = 1}^{n-1} \frac{2 r_0}{r_0^2 +
y_\alpha^2} = - \frac{H'_0 \widetilde{U}}{q} .
\een
%%%
For the $\ud t$ coefficient of the unpaired vielbein, we also need to
use the partial fraction decomposition
%%%
\ben
\frac{\widetilde{U}}{\prod_{k=1}^n (r_0^2 + a_k^2)} =
\sum_{i=1}^n \frac{a_i \gamma_i}{\epsilon_i (r_0^2 + a_i^2)} ,
\een
%%%
hence completing the verification.  It again follows that the Cardy
formulae are satisfied.  The central charges are
%%%
\ben
c_i = \frac{3 \mathcal{A}_{D-2} a_i \prod_{j \neq i} (r_0^2 + a_j^2)
[\prod_{k=1}^{n-1} (r_0^2 + a_k^2) + q r_0^2]}{
\pi V r_0 (r_0^2 + a_i^2)} .
\een

\section{Conclusions}

        In this paper, we have generalized the recently proposed
extremal black hole/CFT correspondence to large classes of charged
rotating black holes in a variety of dimensions.  For extremal black
holes, the near-horizon geometry can be obtained by a limiting (or
decoupling) procedure that implies that the near-horizon geometry is a
solution in its own right.  We started with a general argument that
the near-horizon geometry of extremal rotating black holes is of the
form of a sphere bundle over AdS$_2$, with the connection potentials
proportional to the inverse of the Frolov--Thorne temperatures.  It is
then straightforward to demonstrate that the Cardy formulae for these
near-horizon geometries are satisfied, which we have verified in low
dimensions.  Since the formulae do not rely on any special features of
a particular dimension, they are very likely to be satisfied in
arbitrary dimension.  With this general argument, to show that the
Cardy formulae are satisfied for a particular black hole solution, it
suffices to show that its near-horizon geometry may be cast in a
canonical form.

We then obtained the near-horizon geometries for a variety of charged
rotating black holes in gauged and ungauged supergravities in a
variety of dimensions, and in gravity theories that are low-energy
effective actions of bosonic strings in arbitrary dimension.  In all
of these examples, the near-horizon geometry has the form established
in the general argument.  Consequently the Cardy formulae are
satisfied and the microscopic entropies of the dual CFTs agree with
the Bekenstein--Hawking entropies of the extremal rotating black
holes.

\newpage
\section*{Acknowledgements}

The research of D.D.K.C., H.L. and C.N.P. is supported in part by DOE
grant DE-FG03-95ER40917; M.C. is supported in part by DOE grant
DOE-EY-76-02-3071 and the Fay R. and Eugene L. Langberg Endowed Chair.
M.C. is grateful to the George and Cynthia Woods Mitchell Institute
for Fundamental Physics and Astronomy at Texas A\&M University for
hospitality during the course of this work.

\end{document}